\documentclass[a4paper,fleqn]{cas-dc}
\usepackage[authoryear,compress]{natbib}

\newcommand{\DO}{\textrm{Oxy}}

\begin{document}
\let\WriteBookmarks\relax
\def\floatpagepagefraction{1}
\def\textpagefraction{.001}

\title [mode = title]{Disentangling peri-urban river hypoxia}
%Oxygen-related processes and their spatial variability in an urban river}     
%or Disentangling oxygen-related processes in a highly modified urban river

%\shorttitle{Urban river oxygen processes}
\shorttitle{Peri-urban river hypoxia}
\shortauthors{García-Oliva et~al.}

\author[1]{Ovidio García-Oliva}[%type=editor,
                        %auid=000,bioid=1,
                        %prefix=Sir,
                        %role=Researcher,
                        orcid=0000-0001-6060-2001]
\cormark[1]
\ead{ovidio.garcia@hereon.de}
%\ead[url]{www.cvr.cc, cvr@sayahna.org}
\credit{Conceptualization, Methodology, Data curation, Formal and statistical analysis, Software, Writing-original draft, Writing--review, and editing}

\affiliation[1]{organization={Helmholtz-Zentrum hereon},
    addressline={Max-Planck-Straße 1}, 
    city={Geesthacht},
    % citysep={}, % Uncomment if no comma needed between city and postcode
    postcode={21502}, 
    country={Germany}}

\author[1]{Carsten Lemmen}[orcid=0000-0003-3483-6036]
\ead{carsten.lemmen@hereon.de}
\credit{Funding acquisition, Conceptualization, Software, Writing-original draft, Writing--review, and editing}

\author[2]{Xiangyu Li}[%
   %role=Co-ordinator,
   %suffix=Jr,
   orcid=0000-0001-6377-049X
   ]
\ead{xiangyu.li@io-warnemuende.de}
\credit{Software, Writing--review, and editing}

\affiliation[2]{organization={Leibniz Institute for Baltic Sea Research Warnemünde},
    addressline={Seestrasse 15}, 
    city={Rostock},
    % citysep={}, % Uncomment if no comma needed between city and postcode
    postcode={18119}, 
    country={Germany}}

\author[1]{Kai Wirtz}[orcid=0000-0002-6972-3878]
\ead{kai.wirtz@hereon.de}
\credit{Funding acquisition, Formal analysis, Writing--review, and editing}

\begin{abstract}
Episodes of low dissolved oxygen concentration---hypoxia---threaten the functioning of and the services provided by aquatic ecosystems, particularly those of urban rivers. 
Here, we disentangle oxygen-related processes in the highly modified Elbe River flowing through the major German city of Hamburg, where low oxygen levels are frequently observed. 
We use a process-based biochemical model that describes particulate and dissolved organic matter, micro-algae, their pathogens, and the key reactions that produce or consume oxygen: photosynthesis, re-aeration, respiration, mineralization, and nitrification. 
The model analysis reveals pronounced spatial variability in the relative importance of these processes. 
Photosynthesis and respiration are more prominent upstream of the city, while  mineralization, nitrification, and re-aeration prevail downstream.
The city, characterized by rapid changes in bathymetry, marks a transitional area: pathogen-related micro-algal lysis may increase organic material, explaining the shift towards heterotrophic processes downstream.
As the primary driver of seasonal changes, the model analysis reveals a differential temperature sensitivity of biochemical rates.
These results may be extrapolated to other urban rivers, and also provide valuable information for estuarine water quality management.
\end{abstract}

% \includegraphics{figs/grabs.pdf}

% \begin{highlights}
%   \item Oxygen processes vary pronouncedly along the Elbe River, with photosynthesis and respiration more prominent upstream and mineralization and nitrification downstream.
%   \item Pathogen-related micro-algal lysis increases organic material in urban areas, shifting biochemical processes downstream.
%   \item Differential temperature sensitivity of biochemical rates drives seasonal changes in oxygen dynamics.
% \end{highlights}

% Keywords seperated by \sep

\maketitle

%%%%%%%%%%%%%%%%%%%%%%%%%%%%%%%%%%%%%%%%%%%%%%%%%%%%%%%%%%%%%%%%%%%%%%%%%%%%%%%
\section{Introduction}

%% Why oxygen is important
Dissolved oxygen is a key indicator of water quality and ecosystem health \citep{EC2006}.
Low oxygen levels (hypoxia) harm aerobic organisms and can trigger undesirable processes like greenhouse gas production and the release of nutrients and toxins from sediments \citep{Bastviken2011, Salk2016}.
Despite the dominance of high oxygenation in---mostly turbulent---rivers, increasing evidence shows a trend to low-oxygen conditions worldwide, but particularly in urban areas \citep{Mallin2006, Zhi2023, Blaszczak2023, Ma2024a}.
This deoxygenation threatens the social, ecological, and biogeochemical functioning of urban rivers, such as fisheries or carbon and nutrient retention \citep{Casas-Ruiz2017, Krause2022,  Lespez2022, Grzyb2023}.

%% From/To where is the oxygen coming/going from?
The riverine oxygen budget is governed by biochemical processes acting as sources (e.g., photosynthesis) or sinks (e.g., organic matter mineralization, respiration, abiotic reactions) of oxygen \citep{Yakushev2013, Holzwarth2018a}.
Air-water oxygen exchange---re-aeration---functions as either a source or sink relative to the temperature-dependent water saturation level \citep{Wanninkhof2014, Carter2021}. 
The relative importance of these processes varies with biochemical rates, which are influenced by flow velocity, nutrient loads, light irradiation, and water temperature \citep{Holzwarth2018a, Fan2025}.
These variables exhibit pronounced seasonal and interannual variability, altering the dominance of each process throughout the year.
But also droughts and floods alter particulate matter loads, impacting water turbidity, light availability, nutrient concentrations, and organic matter concentration \citep{Graham2024, Bernal2025}, and thereby affecting photosynthesis rates.
Water temperature is particularly critical, as warming trends correlate with observed deoxygenation \citep{Blaszczak2023}.
Understanding the drivers of oxygen dynamics requires disentangling the controlling biochemical processes responding to seasonally or intermittently changing river conditions.

\begin{figure*}[tb] %htb
 \centering
 \includegraphics[width=0.99\textwidth]{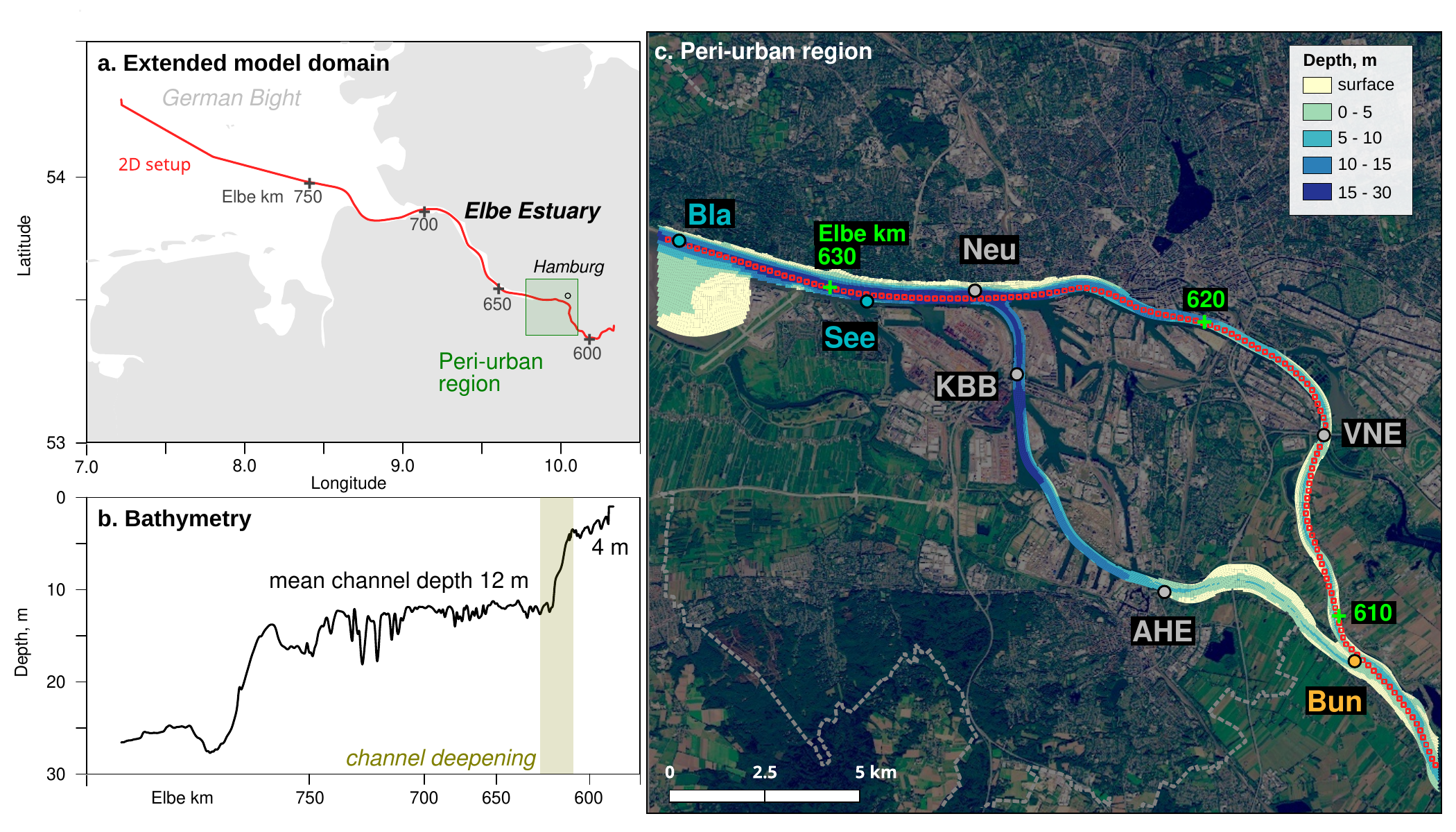} 
 \caption{\textbf{Study area:}
 (a) The horizontal one-dimensional model domain ranges from Geesthacht weir (Elbe km 585) to the central German Bight.  (b) The mean channel depth along this transect defines the bathymetry.
 (c) Focus on the peri-urban region with  sampling stations.
  Upstream sampling stations (in light orange): Zol = Zollenspieker (not shown), and Bun = Bunthaus at Elbe km 599 and 609, respectively.
 Sampling stations in the city (in light gray): AHE = Alte Harburger Elbbrücke, KBB = Köhlbrandbrücke, VNE = Vogelsander Norderelbe, and Neu = Neumühlen at Elbe km 615, 614, 622 and 623, respectively.
 Downstream  stations (in light blue): See = Seemannshöft, and Bla = Blankenese at Elbe km 629 and 634, respectively.
 Copernicus Sentinel data 2025.
 }
 \label{fig:domain}
\end{figure*}

%% Here we paragraph
This study focuses on oxygen-related processes in the highly modified urban Elbe River around Hamburg (Fig.~\ref{fig:domain}), an area frequently experiencing summer hypoxia and fish mortality \citep{Schaffrin2021}.
The Elbe River has a long history of monitoring and modeling efforts \citep[e.g.,][]{Carstens2004, Scbol2014, Pein2021, Pein2025}, with particular attention given to oxygen deficits at the Hamburg port \citep{Schroeder1997, Scbol2014, Hein2018, Fan2025}.
We here employ the biogeochemical model OxyPOM (\textbf{Oxy}gen and \textbf{P}articulate \textbf{O}rganic \textbf{M}atter) coupled to the two-dimensional, depth-resolving river-following hydrodynamical model to simulate oxygen dynamics in 2020-2022.  Our primary goal is the quantification of the relative importance  of biochemical processes driving oxygen dynamics in this urban river in space and time, with emphasis on the role of temperature sensitivities.

%%%%%%%%%%%%%%%%%%%%%%%%%%%%%%%%%%%%%%%%%%%%%%%%%%%%%%%%%%%%%%%%%%%%%%%%%%%%%%%
\section{Methodology}

%%%%%%%%%%%%%%%%%%%%%%%%%%%%%%%%%%%%%%%%%%%%%%%%%%%%%%%%%%%%%%%%%%%%%%%%%%%%%%%
\subsection{OxyPOM model description}

% TODO Needs better figure caption
\begin{figure*}[tb] %htb
 \centering
 \includegraphics[page=1,width=0.99\textwidth]{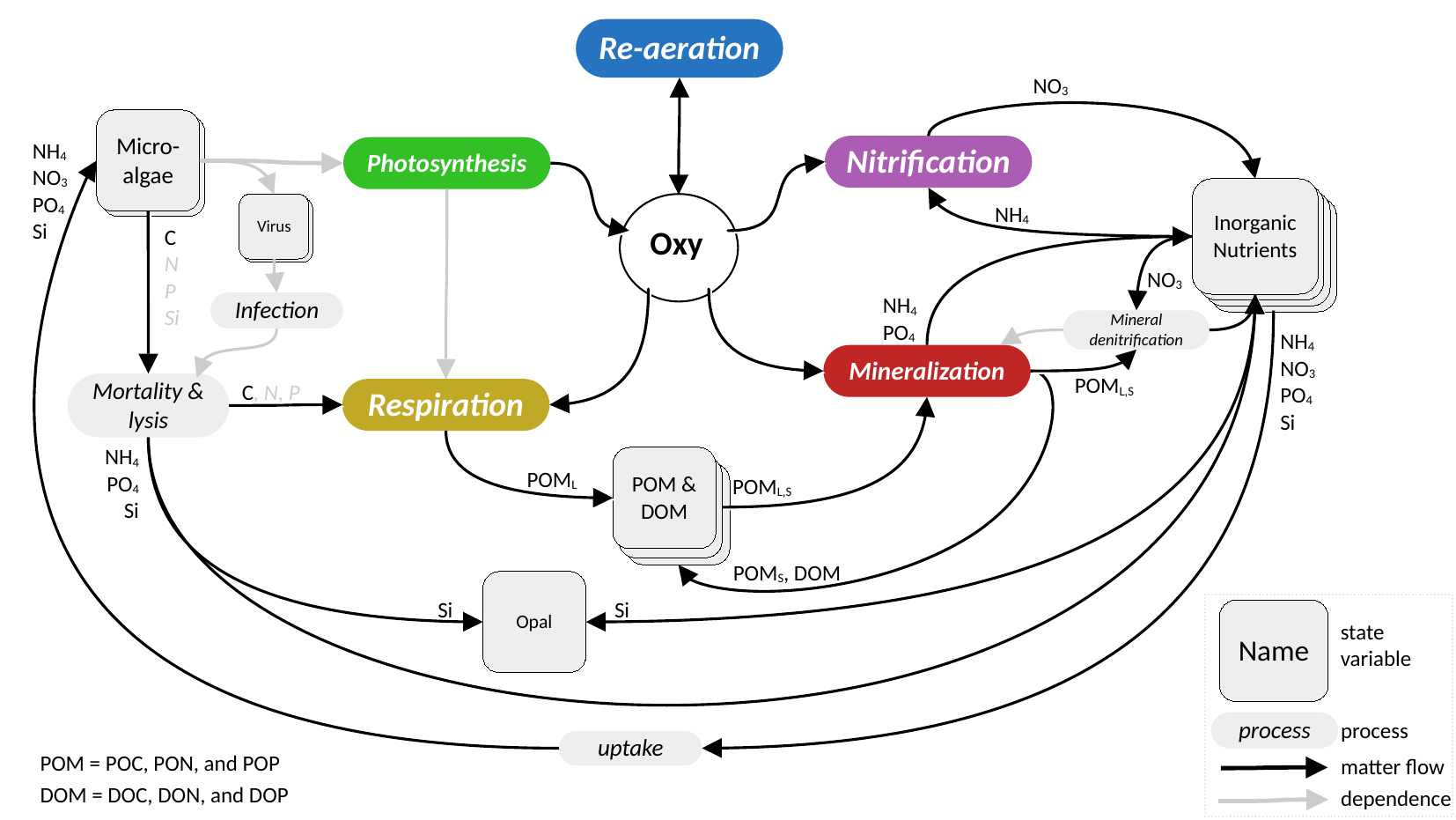} 
 \caption{\textbf{Model configuration:} 
 Processes consuming/producing dissolved oxygen ($\DO$) resolved by OxyPOM.
 Other state variables and fluxes are also shown in a simplified way except for, e.g., settling and light attenuation. 
 Grey symbols denote  implicit fluxes defined by a fixed micro-algae stoichiometry.
 }
 \label{fig:model}
\end{figure*}

A one-dimensional (only horizontal) long-channel predecessor version of OxyPOM was initially implemented by \citet{Holzwarth2018a} in the context of the closed-source Unstructured Tidal Residual Model--Delft Water Quality (UnTRIM DELWAQ) system used by the German Federal Waterways agency. 
This implementation lacked portability and did not adhere to FAIR principles \citep[Findable, Accessible, Interoperable, Reusable][]{Wilkinson2016}.
Our re-implementation as open source utilizes the Framework for Aquatic Biogeochemical Models Application Programming Interface (FABM API) \citep{Bruggeman2014}, thereby ensuring accessibility,  interoperability with other aquatic process models, and reusability across  different hydrodynamic models and zero- to three-dimensional domains.
Beyond the original implementation, we incorporated vertically-explicit formulations for re-aeration in rivers and estuaries \citep{Raymond2001}, primary production, and light attenuation; as new features, we included additional mortality terms for micro-algae, accounting for viral infections \citep{Wirtz2019} and temperature-sensitive loss rates \citep{Scharfe2009}.

%% Basic model description (extremely needed in journals with methods at the end)
OxyPOM resolves the dynamics of dissolved oxygen (\DO), labile and semi-labile particulate organic matter (POM$_\textrm{L,S}$), silicate particles, dissolved organic matter (DOM), dissolved inorganic nutrients, two micro-algae classes, and two pathogenic viruses affecting micro-algae.
POM and DOM are explicitly tracked for carbon, nitrogen, and phosphorus (C, N, P) content, with POM transitioning through labile and semi-labile states before dissolving. 
POM and DOM mineralization releases dissolved inorganic nitrogen, the sum of ammonium and nitrate, and ortho-phosphate; ammonium converts to nitrate depending on oxygen.
Silicate is modeled in dissolved (bio-available) and particulate mineral (opal) forms.
The model resolves two micro-algae classes, one also silicate-dependent representing diatoms. Their growth rates depend on the availability of light, dissolved nitrogen, and of ortho-phosphate.
Micro-algal mortality, which is temperature-dependent and can also result from viral lysis, releases dissolved inorganic nutrients. 

Dissolved oxygen dynamics is based on a mass balance equation:

\begin{equation}\begin{split}
 \frac{\textrm{d} \DO}{\textrm{d}t} = &   \textrm{Re-aeration} + \textrm{Photosynthesis} \\
 &- \textrm{Respiration}  - \textrm{Nitrification} - \textrm{Mineralization}.
 \label{eq:DO}\end{split}
\end{equation}

In Eq.~\ref{eq:DO}, re-aeration (Eq.~\ref{eq:Aer}) at the surface is a function of temperature, salinity, and wind speed \citep{Weiss1970, Wanninkhof1992, Wanninkhof2014}, with a correction for enhanced gas transfer at low wind speeds \citep{Raymond2001}.
Photosynthesis (Eq.~\ref{eq:Pho}) is limited by nutrient concentration and light intensity, following an exponential saturation relationship \citep{Platt1980}.
Respiration (Eq.~\ref{eq:Res}) accounts for oxygen consumption by micro-algae. Oxygen is consumed by
nitrification and mineralization (Eqs.~\ref{eq:Nit} and \ref{eq:Min}, respectively) during ammonia oxidation and organic matter transformation, respectively.
All these processes are expressed as temperature-sensitive biochemical rates using specific $Q_{10}$ coefficients \citep{Sherman2016}.
%[Explain trends]
The full model description is included in the Supplementary material \ref{sm:model-description}.

%%%%%%%%%%%%%%%%%%%%%%%%%%%%%%%%%%%%%%%
\subsection{Study area}

% I miss various important set-up informations:
% how is 1D vertical mixing described?
% is the physics validated? where?
% which physical model is used (GETM? GOTM?)

We applied OxyPOM in the Elbe River using a two-dimensional vertically-resolved along-channel setup (Fig.~\ref{fig:domain}).
In the current application, we used the General Estuarine Transport Model (GETM) \citep{Burchard2002} as a hydrodynamic driver; GETM has demonstrated high skill for the Elbe estuary \citep{Reese2024}.
The horizontal resolution is approximately 300\,m, with a depth-adapting vertical resolution of 15~topography-following $\sigma$-layers--approximately 1\,m at 15\,m depth.
The domain extends from the weir in Geesthacht (stream km 585) to the offshore area of the German Bight, and represents the mean channel depth rather than the maximum depth to maintain appropriate flow velocities \citep{Scbol2014}.
Channel width was approximated by a simple power law formulation \citep{Holzwarth2018a}.
For this analysis, we focused on the freshwater zone of the river section across Hamburg (stream km 600--640), where low oxygen levels are frequently observed \citep{Scbol2014}.

%%%%%%%%%%%%%%%%%%%%%%%%%%%%%%%%%%%%%%%%%%%%%%%%%%%%%%%%%%%%%%%%%%%%%%%%%%%%%%%
\subsection{Data sources}
We reconstructed meteorological conditions, including air temperature, wind velocities, air pressure, total cloud cover, precipitation, and humidity, to an hourly resolution from eight stations operated by the German Weather Service (DWD, \url{https://opendata.dwd.de}).
Boundary conditions on the ocean side were set to mean values for biogeochemical variables at Helgoland Roads \citep{Wiltshire2004a}, with a sinusoidal water height emulating the tidal cycle and a seasonal changing temperature.
On the river side, we reconstructed boundary conditions to a daily resolution from biogeochemical variables observed at the Geesthacht Weir by Wasser- und Schiffahrtsverwaltung (WSV, \url{https://www.kuestendaten.de}) and by Flussgebietsgemeinschaft (FGG) Elbe (\url{https://www.elbe-datenportal.de}), as well as water discharge at the Neu Darchau station from WSV.

For model validation, we used observations of biogeochemical variables---temperature, dissolved oxygen, chlorophyll, ortho-phosphate, nitrate, ammonia, particulate organic carbon (POC), and dissolved organic carbon (DOC)---from eight sampling stations in Hamburg: upstream (Zollenspieker and Bunthaus), in the city (Alte Harburger Elbbrücke, Köhlbrandbrücke, Vogelsander Norderelbe, and Neumühlen), and downstream (Seemannshöft and Blankenese); all data were sourced from WSV and FGG Elbe and downloaded from their official websites. % HOW? web-site? pers. comm.? 
Micro-algae biomass was calculated from chlorophyll concentration using as fixed conversion factor  0.05\,g\,Chl-a g$^{-1}$\,C \citep{Holzwarth2018a}.

%%%%%%%%%%%%%%%%%%%%%%%%%%%%%%%%%%%%%%%%%%%%%%%%%%%%%%%%%%%%%%%%%%%%%%%%%%%%%%%
\subsection{Performance metrics and model validation}
Each model variable was evaluated at each observation station independently. 
We used $N$ pairs of observed $x^\textrm{obs}$ and modelled $x^\textrm{mod}$ values, aggregated as daily means.
We used four different performance metrics.  The (i) coefficient of determination $r^2$ for assessing whether the model captures the observational trend; (ii) the mean bias for assessing over- or underestimation:

\begin{equation}
  \textrm{bias} = \frac{1}{N}\sum_i \left( x^\textrm{mod}_i-x^\textrm{obs}_i \right);
 \label{eq:bias}
\end{equation}

(iii) the normalized square deviation (NSD)---is the model capturing the observed variability?

\begin{equation}
  \textrm{NSD} = \frac{\textrm{SD}(x^\textrm{mod})}{\textrm{SD}(x^\textrm{obs})},
 \label{eq:NSD}
\end{equation}

where $\textrm{SD}(x)=\sqrt{\frac{1}{N-1}\sum_i \left(x_i-\bar{x}\right)^2}$ is the standard deviation with a mean $\bar{x}$; and (iv) the hit rate (HR)---is the model representing the values within a limited range from the observations?

\begin{equation}
  \textrm{HR} = \frac{1}{N}\sum_i n_i,\ \textrm{with}\ n_i = \begin{cases}
    1 & \textrm{if } \left|x^\textrm{mod}_i-x^\textrm{obs}_i\right| \leq D  \\
    0 & \textrm{otherwise} 
  \end{cases},
  \label{eq:HR}
\end{equation}

where $D$ is the deviation, here defined as 20\% of the range of observed values, thus $D=0.2\cdot(\max( x^\textrm{obs}_i)-\min(x^\textrm{obs}_i))$.

%%%%%%%%%%%%%%%%%%%%%%%%%%%%%%%%%%%%%%%%%%%%%%%%%%%%%%%%%%%%%%%%%%%%%%%%%%%%%%%
\section{Results}
%%%%%%%%%%%%%%%%%%%%%%%%%%%%%%%%%%%%%%%%%%%%%%%%%%%%%%%%%%%%%%%%%%%%%%%%%%%%%%%
\subsection{Spatial variability of oxygen sources and sinks}
% Spatial and Temporal Patterns: Patterns observed in the data from the different sampling stations.
\begin{figure*}[tb] %htb
 \centering
 \includegraphics[width=0.99\textwidth]{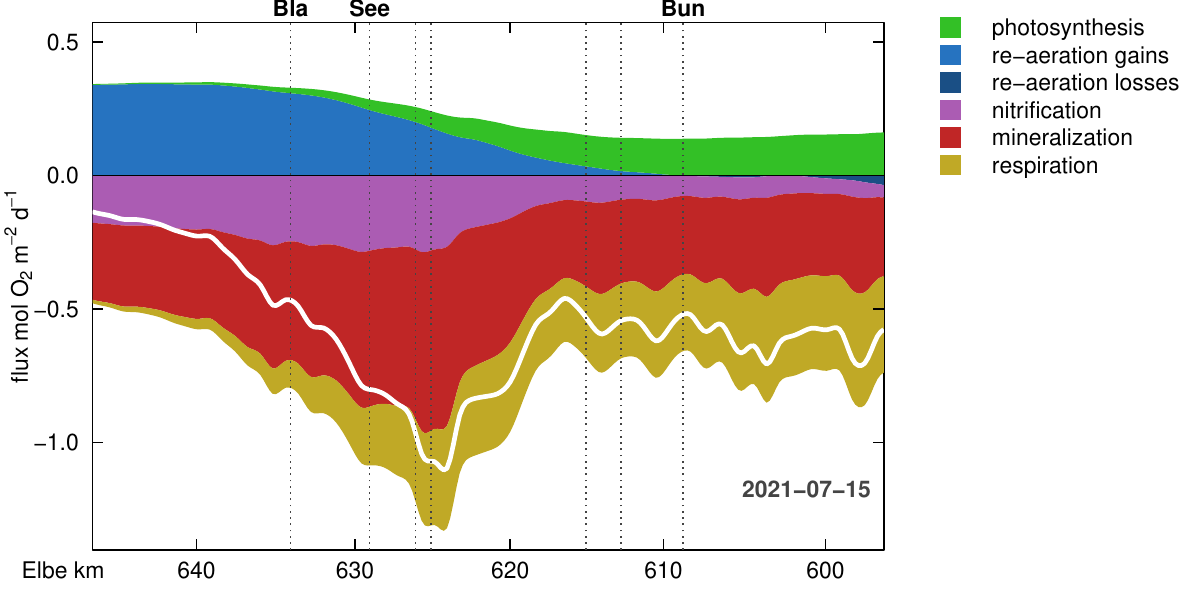} 
 \caption{\textbf{Importance of re-aeration, photosynthesis, nitrification, mineralization and respiration in the oxygen flux (Eq.~\ref{eq:flux}, vertically integrated oxygen dynamics Eq.~\ref{eq:DO}).} 
 The dominant processes driving dissolved oxygen changes along the main channel during the most intense deoxygenation event July 15, 2021. The white line is the sum of all processes, showing a net deoxygenation rate. Bun = Bunthaus, See = Seemannshöft, Bla = Blankenese stations.
 }
 \label{fig:oxygen_flux}
\end{figure*}

%% Spatial variability, description of Fig. 3a
There is a notable degree of spatial variability in the dominance of various biochemical processes within the urban river ecosystem. 
These differences are stronger during summer, when the oxygen minimum occurs (Fig.~\ref{fig:oxygen_flux}). 
Upstream of the city, photosynthesis and respiration dominate. 
The urban environment itself, marked by rapid fluctuations in bathymetry, functions as a transitional zone that alters the dynamics of these processes.
Downstream areas exhibit a greater influence from mineralization, nitrification and re-aeration, indicating a shift in the dominant biochemical activities and physical processes as one moves through the urban landscape.

%%%%%%%%%%%%%%%%%%%%%%%%%%%%%%%%%%%%%%%%%%%%%%%%%%%%%%%%%%%%%%%%%%%%%%%%%%%%%%%
\subsection{Seasonal trends in the oxygen budget}

\begin{figure*}[htb] %htb
 \centering
 \includegraphics[width=0.99\textwidth]{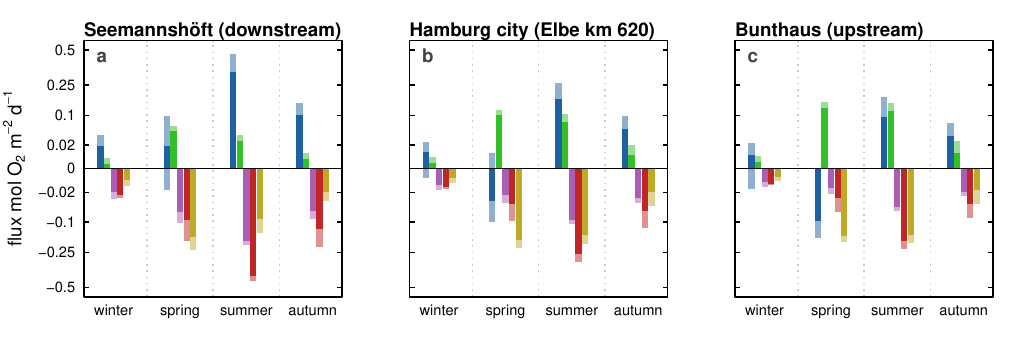} 
 \caption{\textbf{Temporal variation in oxygen dynamics.} 
 The absolute contribution of each process depends on the season as shown for three locations on the river: (a) downstream the city at Seemannshöft, (b) in the city (Elbe km 620), and (c) upstream of the city at Bunthaus. 
 Legend as in Fig.~\ref{fig:oxygen_flux}, dark colors for the median value, lighter colors for the 25--75-percentile interval. 
 Note the square-root-transformed y-axis.
 }
 \label{fig:temporal}
\end{figure*}

%% temporal variability of the relative importance of each process, description of Fig. 3b-d
The leading (de)oxygenation process varies by season (Fig. \ref{fig:temporal}), notably through shifts in the role of re-aeration.
In spring, as photosynthesis increases and becomes the main oxygen source, re-aeration is an oxygen sink, an effect that is more pronounced upstream than in the inner-city and downstream.
Mineralization gains importance in summer and autumn, while respiration is the dominant oxygen sink from late winter into summer, with stronger impacts in the city and downstream.
Nitrification strongly affects the oxygen budget upstream, depending on ammonia concentration, but its downstream contribution appears independent of local sources (see additional figure).
In summer, photosynthesis remains dominant, re-aeration again becomes a source of oxygen, and higher oxygen demand from mineralization and respiration—--driven by elevated organic material and faster micro-algae turnover--—intensify deoxygenation; nitrification also contributes more prominently.
In autumn, photosynthesis still supplies oxygen upstream but not in the city or downstream, and the roles of mineralization, respiration, and nitrification are similar to summer.
In winter, photosynthesis is a source upstream but not downstream, and nitrification becomes the main deoxygenation process.  This observation is consistent with the elevated ammonia discharge in late 2021 that was not observed in other years (see additional figure), alongside comparable respiration and reduced mineralization relative to autumn.

Overall, the net oxygen flux exhibits strong seasonal and along-river variability, with higher deoxygenation rates in summer and autumn (0.3--1.0\,mol\,O$_2$ m$^{-2}$\,d$^{-1}$) and toward downstream locations (cf.\ Fig.~\ref{fig:oxygen_flux}f,g vs.\ \ref{fig:oxygen_flux}e), and weaker deoxygenation in spring (0.3\,mol\,O$_2$ m$^{-2}$\,d$^{-1}$) and near zero in winter.

%%%%%%%%%%%%%%%%%%%%%%%%%%%%%%%%%%%%%%%%%%%%%%%%%%%%%%%%%%%%%%%%%%%%%%%%%%%%%%%
\subsection{Model performance}
\begin{figure*}[b] %htb
 \centering
 \includegraphics[width=0.99\textwidth]{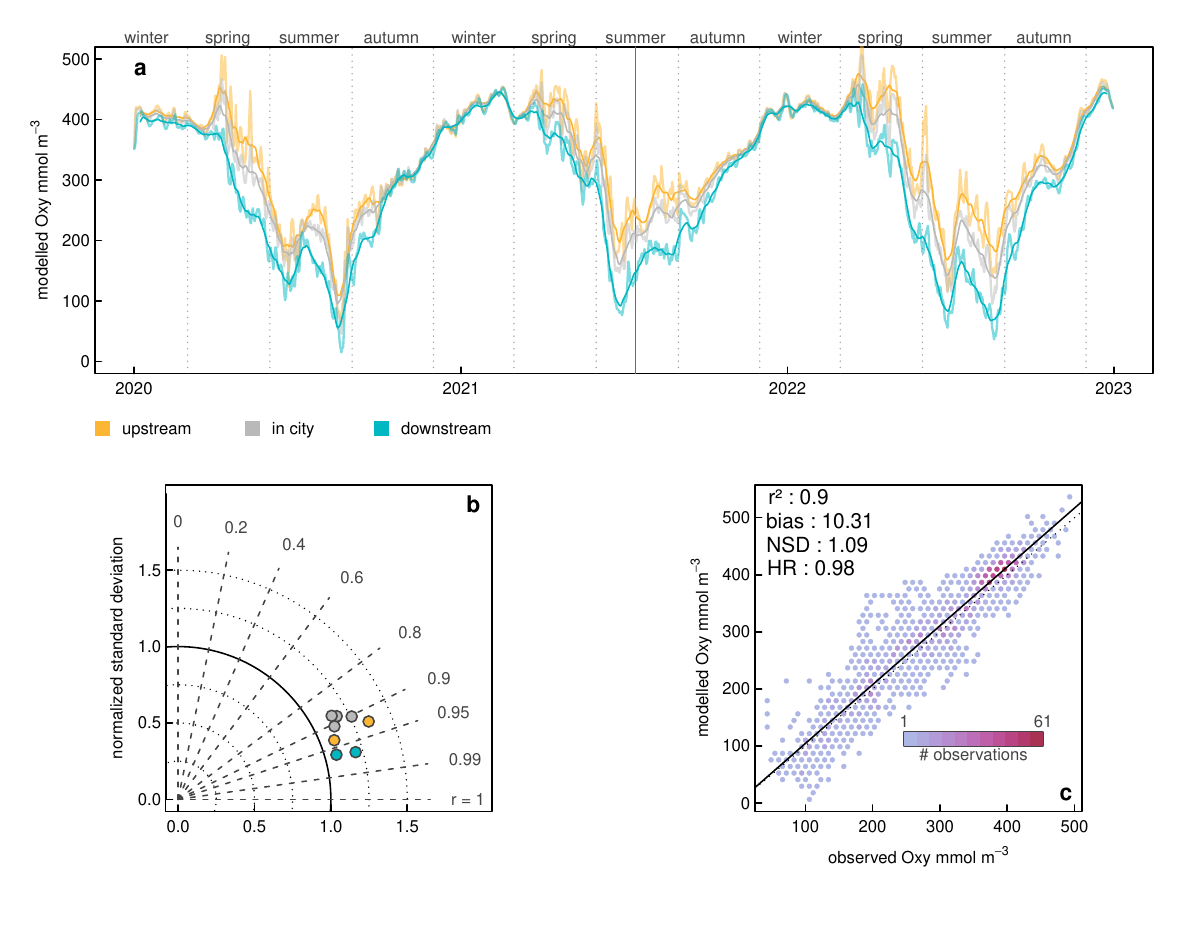} 
 \caption{\textbf{Modelled dissolved oxygen concentration.} 
 (a)~Modelled oxygen upstream in city and downstream (daily mean and 15-day rolling mean in light and dark colors, respectively).
 (b)~Taylor diagram showing the eight stations independently, and (c) comparison of modelled and observed oxygen values. 
 For a complete picture of the oxygen validation, see Fig.~\ref{fig:validation-DO}.
 }
 \label{fig:DO}
\end{figure*}

OxyPOM captures different responses for oxygen in stations upstream, in the city, and downstream  (Fig.~\ref{fig:DO}a).
Larger differences among locations are observed in summer (for reference, red line in Fig.~\ref{fig:DO}a is 15th of July, shown in Fig.~\ref{fig:oxygen_flux}). 
The model has good skill in reproducing oxygen when evaluated for each station independently, with $1.0 < \textrm{NSD} < 1.3$  and $0.85 < r^2 < 0.99$ for all stations (Fig.~\ref{fig:DO}b).
The aggregated evaluation shows good performance, with $r^2 = 0.9$ ($p<2\cdot 10^{-16}$, 2678 degrees of freedom)---the model captures the observed trends, a mean bias of 10.31\,mmol\,m\textsuperscript{-3}---the model overestimate dissolved oxygen, a normalized standard deviation of 1.09---the model reproduces the observed data variability,
 and a hit rate of 0.98 using a deviation of $D$ of 91\,mmol\,m\textsuperscript{-3}---the 98\% of the modelled values lie within a $\pm{}91$\,mmol m\textsuperscript{-3} from the observation (Fig.~\ref{fig:DO}c).
For a extended model validation see the Supplementary material \ref{sec:sm-validation}.

%%%%%%%%%%%%%%%%%%%%%%%%%%%%%%%%%%%%%%%%%%%%%%%%%%%%%%%%%%%%%%%%%%%%%%%%%%%%%%%
%%%%%%%%%%%%%%%%%%%%%%%%%%%%%%%%%%%%%%%%%%%%%%%%%%%%%%%%%%%%%%%%%%%%%%%%%%%%%%%
\section{Discussion}

%% Spatial variability of oxygen dynamics occur for other processes
% Rivers are inhomogenous 
%CHECK THIS... MAKE A PHOTOSYNTHESIS:(RESPIRATION+MINERALIZATION) PLOT
Rivers are often seen as large reactors that process material before it is discharged into the sea \citep{Casas-Ruiz2017}.
Our study demonstrates that the Elbe river is not one homogenous reactor but subdivided in time and space in terms of dominant processes governing oxygen dynamics.
Change points in oxygen dynamics are matched by those in other biogeochemical phenomena.
Phytoplankton abundance, for example, quickly drops downstream of the city \citep{Wiltshire1994, Martens2024}, along with changes in taxa composition \citep{Martens2024a}.
These changes are translated to higher trophic levels, due to modifications in potential food quality and the trophic position of zooplankton \citep{Biederbick2024}.
The city is also a spatial boundary for marine bacteria \citep{Benjamin2024}, silica release \citep{Amann2014}, nitrification and isotopic composition \citep{Sanders2018}, NO production \citep{Ingeniero2024}, and a transitional area for fauna composition \citep{Schwentner2021a}.
Downstream of the city---in the so-called ``oxygen hole'' \citep{Tobias-hunefeldt2024}---fish metabolites reveal respiration stress due to low oxygen conditions \citep{Koll2024a}, and massive fish mortalities have been reported by news media, such as for 2021 (\url{https://www.wwf.de/2021/juni/grosses-sauerstoffloch-in-der-tideelbe-erste-hinweise-auf-fischsterben}).
% TODO: reference newspaper article correctly, it does not show in the print version
Moreover, the river's function changes while crossing the city:
Upstream, the river can be characterized as autotrophic due to higher photosynthesis rates with net oxygen production and carbon fixation; downstream, the river becomes heterotrophic due to higher mineralization rates with net oxygen consumption and carbon emission \citep{Rewrie2025}.
A similar heterogeneity in leading (de)oxygenation processes has been observed in other urban rivers such as [here name some examples] \citep{Huang2017a, He2024}.

% Our results show a strong seasonal variability in the processes explaining oxygen dynamics in relative and absolute terms.
% Photosynthesis has been thought to be relevant only in spring and summer as micro-algae loads during autumn and winter are relatively low.
% Here, however, we show a that decent photosynthesis rates are possible through the year, specially upstream, where light conditions, despite of high turbidity of the winter months, can be still more favourable than in deeper downstream and transitional sections of the river.
% This is supported by observations of year-round presence of picoplankton \citep{Martens2024}.
% Nevertheless, micro-algae presence is hundreds-fold higher in spring-autumn, when higher turnover rates are possible due to higher temperatures, especially in the better light conditions of the shallow upstream area.
% It is there where POM is produced and then transported downstream where it will consume oxygen during their mineralization.  

%% main oxygenation processes
As re-aeration depends on oxygen levels, its role may shift from a source to a sink depending on whether the water is saturated \citep{Wanninkhof2014}. 
As a surface process, re-aeration requires complete mixing of the water column to be effective as an oxygen source; otherwise, it primarily occurs in shallow waters, leaving deeper areas with low oxygen. 
Consequently, alterations in vertical mixing and stratification are closely connected to oxygen dynamics \citep{Fan2025}.
However, this has led to misconceptions about the relative importance of biochemical processes to explain hypoxia and anoxia in the Elbe River. 

%%%%%%%%%%%%%%%%%%%%%%%%%%%%%%%%%%%%%%%%%%%%%%%%%%%%%%%%%%%%%%%%%%%%%%%%%%%%%%%
\subsection{The role of temperature in oxygen dynamics}
The role of temperature in explaining oxygen deficits has often been limited to its physical effects: first, by reducing oxygen solubility in warmer water \citep{Weiss1970}, and second, by increasing stratification that isolates deep water, preventing efficient re-aeration \citep{Pein2021, Fan2025, VanVliet2023}.
Nevertheless, temperature-sensitivity in biochemical processes may have a stronger influence on the oxygen budget than previously accounted for. 
As water warms, oxygen-related processes become unbalanced due to differential temperature sensitivity in grazing \citep{Verity2002, Ferreira2022}, micro-algae growth rates \citep{Sherman2016, Anderson2021}, photosynthesis--light responses \citep{Coles2000}, viral infections \citep{Padhy1977}, and other chemical rates \citep{Garnier1999}. 
Our model-based analysis emphasizes that the differential sensitivity of biochemical rates to temperature changes is a key factor driving seasonal variations in (de)oxygenation.
As temperature fluctuates throughout the year, the rates of photosynthesis, respiration, mineralization, and nitrification respond differently, leading to shifts in overall ecosystem dynamics. 
These effects are not always accounted for in other model applications that assume the same sensitivities across all biochemical processes \citep[e.g.,][]{Huang2017a}.

%HERE IS TO DISCUSS THE MISSING RESULTS

Our results suggest that the account for he differential temperature sensitivity in oxygen-related biochemical processes is key for evaluating the effects of climate warming \citep{Hein2018} or transitory extreme events such as heatwaves in oxygen dynamics \citep{VanVliet2023, Lyu2025}.
Most models implement a uniform temperature sensitivity---all processes are affected by temperature changes in the same magnitude, thus creating responses that vary in magnitude but not in the relative importance of each process \citep{Huang2017a}.
This may be problematic for identifying the processes explaining oxygen dynamics.
Also other  processes found relevant in our study are usually absent in models: for example, pathogens---here  viral infection, is systematically neglected in in models with explicit oxygen dynamics \citep[e.g.,][]{Yakushev2013, Neumann2022}.  Viral lysis is a potentially important mortality term for micro-algae \citep{Wirtz2019, Elovaara2020, Krishna2024} and an important source of organic material that fuels high mineralization rates downstream.
%%Other toppics to discuss Gas emissions (NxOy, CH4, CO2, H2S, etc.)
%%loses of ecosystem services
%%close with the loss of the ``cleaning power'' of the river reactor due to low oxygen.

%%%%%%%%%%%%%%%%%%%%%%%%%%%%%%%%%%%%%%%%%%%%%%%%%%%%%%%%%%%%%%%%%%%%%%%%%%%%%%%%
\subsection{Changes in oxygen co-occur with other leading biochemical processes}
In rivers, the oxygen dynamics shape other biochemical processes through positive or negative feedbacks \citep{Carter2021, Lewis2024}. 
For example, denitrification---the mineralization of organic material using the oxygen present in NO\textsubscript{3}, which is subsequently reduced to N\textsubscript{2}O and N\textsubscript{2}---increases under low oxygen conditions and indirectly influences  the oxygen budget in two ways\citep{Zeng2019b}. First, 
denitrification can positively affect the oxygen budget due to its contribution to organic matter mineralization, thus decreasing oxygen consumption.
Secondly, organic matter mineralization produces ammonia, which increases oxygen demand due to nitrification \citep{Cai2022}.
In a river, this oxygen consumption may occur downstream, explaining the prominence of nitrification after the city (Fig. \ref{fig:oxygen_flux}a).

The potent greenhouse case N$_2$O is an intermediate product during both denitrification and nitrification \citep{Brase2017}, with nitrification being the dominant source in the Elbe River.
Its production and destruction are  tightly linked to oxygen dynamics: lower oxygen is associated with higher N$_2$O, while under anoxic conditions N$_2$O is further reduced by denitrification \citep{Stenstroem2014}.
The amount of nitrous oxide produced during nitrification and denitrification can vary widely, making it difficult to estimate its emissions from rivers.
Estimates or measurements of N$_2$O yields range from 0.01 to 1\%; \citet{Beaulieu2011} found a median value of less than 1\% across 72 streams in the United States; \citet{deWilde2000} found factors between 0.1 and 0.4\% for nitrification in the Scheldt estuary. 
In the Elbe, \citet{Ingeniero2024} observed usually less than 0.1\% of dissolved inorganic nitrogen in the form of N$_2$O, but elevated levels (up to 40\,nmol\,L$^{-1}$) were found in the Hamburg port area.
\citet{Brase2017} had previously observed up to 52\,nmol\,L$^{-1}$ in this area.
While the river is usually saturated with N$_2$O, it is oversaturated in summer up to 200\% and therefore a source of N$_2$O to the atmosphere with emissions on the order of 20\,$\mu$mol\,m$^{-2}$\,d$^{-1}$ \citep{Schulz2023}.
Intermittently and spatially restricted to the city area, supersaturation has been observed up to 400\% \citep{Brase2017}, with emissions around 100\,$\mu$mol\,m$^{-2}$\,d$^{-1}$ \citep{Schulz2023}.
\citet{Preussler2025} was able to represent this value in a preliminary modeling study using OxyPOM.

Similar dynamics to those of (de-)nitrification may occur with other processes whose effects on oxygen dynamics are unclear--or at least not trivial. 
For example, while viral lysis can be viewed just as an additional mortality term for micro-algae, diseases may have complex feedback interactions with nutrient availability \citep{Krishna2024, Thongthaisong2025}, further impacting oxygen dynamics through increased photosynthesis rates.
This process might be particularly important under low discharge and high irradiance conditions, where upstream water is nutrient-depleted \citep{Kamjunke2021, Schulz2023}, and photosynthesis may rely on local nutrient recycling.
However, these positive effects are limited, as viral lysis increases labile POM, which rapidly consumes oxygen during mineralization and decreases light availability due to shading.
Furthermore, local nutrient recycling is conditioned on high light availability, which partly explains low  phytoplankton growth rate downstream of the city, where changes in channel depth and smaller flow velocities lead to POM accumulation  and low light availability \citep{Kerner2007}.

% SOMETHING MORE TO SAY? No, enough tsaid

%%%%%%%%%%%%%%%%%%%%%%%%%%%%%%%%%%%%%%%%%%%%%%%%%%%%%%%%%%%%%%%%%%%%%%%%%%%%%%%%
\subsection{Missing bits may (not) matter}
Our physical-biogeochemical model neglects a series of biological, chemical and physical processes linking oxygen and biochemical dynamics.
%Some of them may matter, some others may just add complexity without adding skill.
Alike almost all biogeochemical models, OxyPOM ignores higher trophic levels, such as zooplankton, fish and marine mammals, which not only determine river ecology \citep{VanNeer2023, Biederbick2024}, but may cascade down to lower trophic levels and biogeochemistry \citep{Pace1999, Eger2020}.
In OxyPOM, more complex predator dynamics is only expressed as a temperature-dependent mortality rate so that the model may miss relevant biological processes directly affecting dissolved oxygen.
For example, when fish die due to low oxygen, the grazing pressure on the predators of micro-algae may be reduced, creating a sudden high predation pressure on micro-algae \citep{Rettig2021} and, in consequence, reducing photosynthetic oxygen production. This together with increased organic matter degradation due to deceasing fish likely spurs a positive feedback loop .
Therefore, the influence of trophic cascades during massive fish mortalities may define an important research gap also relevant for water quality and fisheries management.

Our model prescribes micro-algae with a fixed stoichiometry following the view that rivers are not nutrient limited and, due to a high turbidity, light is the limiting resource, as observed in the Elbe River \citep{Rewrie2025}.
This reasonable simplification, however, may not be always valid, as water upstream can occasionally become nutrient-depleted \citep{Kamjunke2021}, especially in the future due to de-eutrophication efforts and changes in the river discharge \citep{Schulz2023}.
%Light acclimation and flexible chlorophyll content is also missing, 
%CAN I EXPLAIN MISSMATCHES IN ALG USING THIS?
% I would not elaborate as this seciton is already long enough

Although mineralization and nitrification are effectively the main biochemical pathways to oxygen consumption, other chemical reactions, such as SO\textsubscript{4} formation, can be also significant \citep{Spieckermann2022} in our study area.
These reactions, together with others redox processes---Fe and Mn oxidation---are necessary to explain oxygen dynamics in hypoxic and anoxic waters and have been already considered in other models \citep{Yakushev2017}. 
Especially, the continuous oxygen demand by the sediments is not explicitly resolved here. 
Such a missing oxygen sink may also be responsible for the the slight overestimation of oxygen levels in the simulations .
%Other chemical processes as partial nitrification--production of nitrite NO\textsubscript{2}-- are not included
%EXPAND THIS? Rather not

%Heavy metals mobilization due to low oxygen and possible feedback.
%Metals from upstream and the role of oxygen \citep{Rinklebe2019}
%EXPAND THIS? 

Finally, overlooked physical feedbacks may also impact (de)oxygenation, as for example in the role of suspended particulate matter for light availability.
Suspended particulate matter is abundant in the city section of the river \citep{Kerner2007}, with the potential of changing light availability and enhancing vertical transport of POM in tidal affected environment \citep{Li2025}.
This process, while critical to understand light limitation is overly simplified in our model.
We assumed a constant concentration of inorganic suspended particulate matter through the year, while particle density depends on discharge \citep{Baborowski2012}, season \citep{Tobias-hunefeldt2024}, and erosion intensity (after events) \citep{Uber2022}.

Additional shortcomings may originate from the physical limitations of a simplified topology.
For example, our simple one-channel funnel-like setup ignores much of the complexity on the water channels in the city. 
More realistic and numerically demanding modelling applications, find that these features can become critical for oxygen dynamics and phytoplankton persistence \citep{Fan2025, Steidle2024}. 

Despite all these limitations, our model configuration achieves high levels of accuracy comparable with other studies \citep{Holzwarth2018a, Pein2021, Fan2025} or even better \citep{Schroeder1997, Scbol2014, Hein2018}.
A two-dimensional vertically resolved long-channel topology describes the oxygen related processes in the peri-urban area roughly in par with a three-dimensional set-up;  adding a vertical dimension to a prior one-dimensional topology introduces important physicochemical feedbacks---e.g., stratification and re-aeration.
Our model approach especially reproduces low oxygen levels, a difficult task for other models \citep{Scbol2014, Hein2018}.
We could attribute this in part to (1) the differential temperature-dependencies  of the biochemical and ecological processes, and (2) a an improved representations of light limitation and re-aeration..

Despite all these simplifications, limitations and omissions in biological, chemical and physical processes, our model configuration achieves high levels of accuracy comparable with other studies \citep{Pein2021, Fan2025, Holzwarth2018a} or even better \citep{Scbol2014, Hein2018, Schroeder1997}.
A two-dimensional vertically resolved long-channel topology's performance for describing the oxygen processes in the peri-urban area is on par with a three-dimensional one;  adding a vertical dimension to a prior one-dimensional topology introduces important physicochemical feedbacks---e.g., stratification and re-aeration.
OxyPOM especially excels in reproducing low oxygen levels, a difficult task for other models \citep{Scbol2014, Hein2018}.
We attribute this to first, including the relevant processes the temperature-dependent nature of the biochemical process, and second, a vertical dimension with improved representations for light limitation and re-aeration, as well as of other sources of oxygen.

\paragraph{Data availability.}
All data was obtained from 
DWD \url{https://opendata.dwd.de} and used under CC BY 4.0 license(\url{https://www.dwd.de/copyright}); 
WSV (\url{https://www.kuestendaten.de}) and used under DL-DE->Zero-2.0 license (\url{https://www.govdata.de/dl-de/zero-2-0}); and 
FGG Elbe \url{https://www.elbe-datenportal.de} used under the terms of their proprietary license.
Scripts to download third-party data required to reproduce analyses are freely available at \url{https://github.com/ovgarol/elbe-oxygen} and archived in Zenodo \citep{Garcia-Oliva2025-oxypom}. 

The results of our simulations will be available at \url{https://pangaea.de/}. 

\paragraph{Code availability.}
The source code for OxyPOM, GETM, and the Hamburg 2D setup are available at \url{https://codebase.helmholtz.cloud/dam-elbextreme/oxypom}, at \url{https://sourceforge.net/p/getm/code/ci/iow/tree}, and \url{https://codebase.helmholtz.cloud/dam-elbextreme/getm-setup-elbe-2d}, respectively.  The source code for OxyPOM is also archived in Zenodo \citep{Garcia-Oliva2025-oxypom}.
Custom scripts were used for simulation studies and data analyses. 
All code to generate figures is publicly available. 
All code is publicly available at \url{https://github.com/ovgarol/elbe-oxygen} and archived in Zenodo \citep{Garcia-Oliva2025-oxypom}.

\paragraph{Acknowledgments.} 
This study was made possible by grants No. 03F0954D and No. 03F0954F of the German Federal Ministry of Research, Technology and Space (BMFTR) as part of the DAM mission ``mareXtreme'', project ``ElbeXtreme'';  it was supported by the Helmholtz Association with their Innovation Pool for the Research Field Earth and Environment AGRIO: Effect of anthropogenic modifications and climate change on greenhouse gas emissions along the river-ocean continuum and, and  through the joint research program ``Changing Earth - Sustaining our Future''.
This work used resources of the Deutsches Klimarechenzentrum (DKRZ) granted by its Scientific Steering Committee under project ID ``gg0877''.

\paragraph{Competing Interests Statement.}
We declare to have no competing interests.

\printcredits

\bibliographystyle{cas-model2-names}

%% This is to keep a bib file just with used references
%% bibexport -o oxypom.bib cas-dc-template.aux
%% remember to change to oxypom.bib after running
\bibliography{oxypom.bib}

\begin{thebibliography}{87}
\expandafter\ifx\csname natexlab\endcsname\relax\def\natexlab#1{#1}\fi
\providecommand{\url}[1]{\texttt{#1}}
\providecommand{\href}[2]{#2}
\providecommand{\path}[1]{#1}
\providecommand{\DOIprefix}{doi:}
\providecommand{\ArXivprefix}{arXiv:}
\providecommand{\URLprefix}{URL: }
\providecommand{\Pubmedprefix}{pmid:}
\providecommand{\doi}[1]{\href{http://dx.doi.org/#1}{\path{#1}}}
\providecommand{\Pubmed}[1]{\href{pmid:#1}{\path{#1}}}
\providecommand{\bibinfo}[2]{#2}
\ifx\xfnm\relax \def\xfnm[#1]{\unskip,\space#1}\fi
%Type = Article
\bibitem[{Amann et~al.(2014)Amann, Weiss and Hartmann}]{Amann2014}
\bibinfo{author}{Amann, T.}, \bibinfo{author}{Weiss, A.},
  \bibinfo{author}{Hartmann, J.}, \bibinfo{year}{2014}.
\newblock \bibinfo{title}{{Silica fluxes in the inner Elbe Estuary, Germany}}.
\newblock \bibinfo{journal}{Biogeochemistry} \bibinfo{volume}{118},
  \bibinfo{pages}{389--412}.
\newblock \DOIprefix\doi{10.1007/s10533-013-9940-3}.
%Type = Article
\bibitem[{Anderson et~al.(2021)Anderson, Barton, Clayton, Dutkiewicz and
  Rynearson}]{Anderson2021}
\bibinfo{author}{Anderson, S.I.}, \bibinfo{author}{Barton, A.D.},
  \bibinfo{author}{Clayton, S.}, \bibinfo{author}{Dutkiewicz, S.},
  \bibinfo{author}{Rynearson, T.A.}, \bibinfo{year}{2021}.
\newblock \bibinfo{title}{{Marine phytoplankton functional types exhibit
  diverse responses to thermal change}}.
\newblock \bibinfo{journal}{Nat. Commun.} \bibinfo{volume}{12},
  \bibinfo{pages}{1--9}.
\newblock \DOIprefix\doi{10.1038/s41467-021-26651-8}.
%Type = Article
\bibitem[{Ault(2023)}]{Ault2023}
\bibinfo{author}{Ault, J.E.}, \bibinfo{year}{2023}.
\newblock \bibinfo{title}{{A River Runs through It: The Elbe, Socialist
  Security, and East Germany's Borders}}.
\newblock \bibinfo{journal}{Cent. Eur. Hist.} \bibinfo{volume}{56},
  \bibinfo{pages}{196--213}.
\newblock \DOIprefix\doi{10.1017/S0008938922001030}.
%Type = Article
\bibitem[{Baborowski et~al.(2012)Baborowski, Simeonov and
  Einax}]{Baborowski2012}
\bibinfo{author}{Baborowski, M.}, \bibinfo{author}{Simeonov, V.},
  \bibinfo{author}{Einax, J.W.}, \bibinfo{year}{2012}.
\newblock \bibinfo{title}{{Assessment of Water Quality in the Elbe River at
  Flood Water Conditions Based on Cluster Analysis, Principle Components
  Analysis, and Source Apportionment}}.
\newblock \bibinfo{journal}{Clean - Soil, Air, Water} \bibinfo{volume}{40},
  \bibinfo{pages}{373--380}.
\newblock \DOIprefix\doi{10.1002/clen.201100085}.
%Type = Article
\bibitem[{Bastviken et~al.(2011)Bastviken, Tranvik, Downing, Crill and
  Enrich-Prast}]{Bastviken2011}
\bibinfo{author}{Bastviken, D.}, \bibinfo{author}{Tranvik, L.J.},
  \bibinfo{author}{Downing, J.A.}, \bibinfo{author}{Crill, P.M.},
  \bibinfo{author}{Enrich-Prast, A.}, \bibinfo{year}{2011}.
\newblock \bibinfo{title}{{Freshwater methane emissions offset the continental
  carbon sink.}}
\newblock \bibinfo{journal}{Science} \bibinfo{volume}{331},
  \bibinfo{pages}{50}.
\newblock \DOIprefix\doi{10.1126/science.1196808}.
%Type = Article
\bibitem[{Beaulieu et~al.(2011)}]{Beaulieu2011}
\bibinfo{author}{Beaulieu, J.J.}, et~al., \bibinfo{year}{2011}.
\newblock \bibinfo{title}{Nitrous oxide emission from denitrification in stream
  and river networks}.
\newblock \bibinfo{journal}{Proceedings of the National Academy of Sciences}
  \bibinfo{volume}{108}, \bibinfo{pages}{214--219}.
\newblock \DOIprefix\doi{10.1073/pnas.1011464108}.
%Type = Article
\bibitem[{Bernal et~al.(2025)Bernal, Ledesma, Pe{\~{n}}arroya, Jativa,
  Catal{\'{a}}n, Casamayor, Lupon, Marc{\'{e}}, Mart{\'{i}},
  Triad{\'{o}}-Margarit and Rocher-Ros}]{Bernal2025}
\bibinfo{author}{Bernal, S.}, \bibinfo{author}{Ledesma, J.L.J.},
  \bibinfo{author}{Pe{\~{n}}arroya, X.}, \bibinfo{author}{Jativa, C.},
  \bibinfo{author}{Catal{\'{a}}n, N.}, \bibinfo{author}{Casamayor, E.O.},
  \bibinfo{author}{Lupon, A.}, \bibinfo{author}{Marc{\'{e}}, R.},
  \bibinfo{author}{Mart{\'{i}}, E.}, \bibinfo{author}{Triad{\'{o}}-Margarit,
  X.}, \bibinfo{author}{Rocher-Ros, G.}, \bibinfo{year}{2025}.
\newblock \bibinfo{title}{{Expanding towards contraction: the alternation of
  floods and droughts as a fundamental component in river ecology}}.
\newblock \bibinfo{journal}{Biogeochemistry} \bibinfo{volume}{168},
  \bibinfo{pages}{11}.
\newblock \DOIprefix\doi{10.1007/s10533-024-01197-1}.
%Type = Article
\bibitem[{Biederbick et~al.(2025)Biederbick, M{\"{o}}llmann, Hauten, Russnak,
  Lahajnar, Hansen, Dierking and Koppelmann}]{Biederbick2024}
\bibinfo{author}{Biederbick, J.}, \bibinfo{author}{M{\"{o}}llmann, C.},
  \bibinfo{author}{Hauten, E.}, \bibinfo{author}{Russnak, V.},
  \bibinfo{author}{Lahajnar, N.}, \bibinfo{author}{Hansen, T.},
  \bibinfo{author}{Dierking, J.}, \bibinfo{author}{Koppelmann, R.},
  \bibinfo{year}{2025}.
\newblock \bibinfo{title}{{Spatial and temporal patterns of zooplankton trophic
  interactions and carbon sources in the eutrophic Elbe estuary (Germany)}}.
\newblock \bibinfo{journal}{ICES J. Mar. Sci.} \bibinfo{volume}{82}.
\newblock \DOIprefix\doi{10.1093/icesjms/fsae189}.
%Type = Article
\bibitem[{Blaszczak et~al.(2023)Blaszczak, Koenig, Mejia, G{\'{o}}mez‐Gener,
  Dutton, Carter, Grimm, Harvey, Helton and Cohen}]{Blaszczak2023}
\bibinfo{author}{Blaszczak, J.R.}, \bibinfo{author}{Koenig, L.E.},
  \bibinfo{author}{Mejia, F.H.}, \bibinfo{author}{G{\'{o}}mez‐Gener, L.},
  \bibinfo{author}{Dutton, C.L.}, \bibinfo{author}{Carter, A.M.},
  \bibinfo{author}{Grimm, N.B.}, \bibinfo{author}{Harvey, J.W.},
  \bibinfo{author}{Helton, A.M.}, \bibinfo{author}{Cohen, M.J.},
  \bibinfo{year}{2023}.
\newblock \bibinfo{title}{{Extent, patterns, and drivers of hypoxia in the
  world's streams and rivers}}.
\newblock \bibinfo{journal}{Limnol. Oceanogr. Lett.} \bibinfo{volume}{8},
  \bibinfo{pages}{453--463}.
\newblock \DOIprefix\doi{10.1002/lol2.10297}.
%Type = Misc
\bibitem[{Branoff et~al.(2024)Branoff, Gr{\"{u}}terich, Wilson,
  Tobias-Hunefeldt, Saadaoui, Mittmann-Goetsch, Neiske, Lexmond, Becker,
  Grossart, Porada, Streit, Eschenbach, Kutzbach and Jensen}]{Benjamin2024}
\bibinfo{author}{Branoff, B.B.}, \bibinfo{author}{Gr{\"{u}}terich, L.},
  \bibinfo{author}{Wilson, M.}, \bibinfo{author}{Tobias-Hunefeldt, S.P.},
  \bibinfo{author}{Saadaoui, Y.}, \bibinfo{author}{Mittmann-Goetsch, J.},
  \bibinfo{author}{Neiske, F.}, \bibinfo{author}{Lexmond, F.},
  \bibinfo{author}{Becker, J.N.}, \bibinfo{author}{Grossart, H.P.},
  \bibinfo{author}{Porada, P.}, \bibinfo{author}{Streit, W.R.},
  \bibinfo{author}{Eschenbach, A.}, \bibinfo{author}{Kutzbach, L.},
  \bibinfo{author}{Jensen, K.}, \bibinfo{year}{2024}.
\newblock \bibinfo{title}{{Partitioning biota along the Elbe River estuary:
  where are the community transitions?}}
%Type = Article
\bibitem[{Brase et~al.(2017)Brase, Bange, Lendt, Sanders and
  D{\"a}hnke}]{Brase2017}
\bibinfo{author}{Brase, L.}, \bibinfo{author}{Bange, H.W.},
  \bibinfo{author}{Lendt, R.}, \bibinfo{author}{Sanders, T.},
  \bibinfo{author}{D{\"a}hnke, K.}, \bibinfo{year}{2017}.
\newblock \bibinfo{title}{High resolution measurements of nitrous oxide (n2o)
  in the elbe estuary}.
\newblock \bibinfo{journal}{Frontiers in Marine Science} \bibinfo{volume}{4},
  \bibinfo{pages}{162}.
\newblock \DOIprefix\doi{10.3389/fmars.2017.00162}.
%Type = Article
\bibitem[{Bruggeman and Bolding(2014)}]{Bruggeman2014}
\bibinfo{author}{Bruggeman, J.}, \bibinfo{author}{Bolding, K.},
  \bibinfo{year}{2014}.
\newblock \bibinfo{title}{{A general framework for aquatic biogeochemical
  models}}.
\newblock \bibinfo{journal}{Environ. Model. Softw.} \bibinfo{volume}{61},
  \bibinfo{pages}{249--265}.
\newblock \DOIprefix\doi{10.1016/j.envsoft.2014.04.002}.
%Type = Article
\bibitem[{Burchard and Bolding(2002)}]{Burchard2002}
\bibinfo{author}{Burchard, H.}, \bibinfo{author}{Bolding, K.},
  \bibinfo{year}{2002}.
\newblock \bibinfo{title}{{GETM – a general estuarine transport model.
  Scientific documentation.}}
\newblock \bibinfo{journal}{Tech. Rep. EUR 20253 en} , \bibinfo{pages}{164}.
%Type = Article
\bibitem[{Cai et~al.(2022)Cai, Hong, Wu, Moore, Vamerali, Ye and
  Wang}]{Cai2022}
\bibinfo{author}{Cai, M.}, \bibinfo{author}{Hong, Y.}, \bibinfo{author}{Wu,
  J.}, \bibinfo{author}{Moore, S.S.}, \bibinfo{author}{Vamerali, T.},
  \bibinfo{author}{Ye, F.}, \bibinfo{author}{Wang, Y.}, \bibinfo{year}{2022}.
\newblock \bibinfo{title}{{Nitrate Addition Increases the Activity of Microbial
  Nitrogen Removal in Freshwater Sediment}}.
\newblock \bibinfo{journal}{Microorganisms} \bibinfo{volume}{10}.
\newblock \DOIprefix\doi{10.3390/microorganisms10071429}.
%Type = Article
\bibitem[{Carstens et~al.(2004)Carstens, Claussen, Bergemann and
  Gaumert}]{Carstens2004}
\bibinfo{author}{Carstens, M.}, \bibinfo{author}{Claussen, U.},
  \bibinfo{author}{Bergemann, M.}, \bibinfo{author}{Gaumert, T.},
  \bibinfo{year}{2004}.
\newblock \bibinfo{title}{{Transitional waters in Germany: The Elbe estuary as
  an example}}.
\newblock \bibinfo{journal}{Aquat. Conserv. Mar. Freshw. Ecosyst.}
  \bibinfo{volume}{14}, \bibinfo{pages}{81--92}.
\newblock \DOIprefix\doi{10.1002/aqc.652}.
%Type = Article
\bibitem[{Carter et~al.(2021)Carter, Blaszczak, Heffernan and
  Bernhardt}]{Carter2021}
\bibinfo{author}{Carter, A.M.}, \bibinfo{author}{Blaszczak, J.R.},
  \bibinfo{author}{Heffernan, J.B.}, \bibinfo{author}{Bernhardt, E.S.},
  \bibinfo{year}{2021}.
\newblock \bibinfo{title}{{Hypoxia dynamics and spatial distribution in a low
  gradient river}}.
\newblock \bibinfo{journal}{Limnol. Oceanogr.} \bibinfo{volume}{66},
  \bibinfo{pages}{2251--2265}.
\newblock \DOIprefix\doi{10.1002/lno.11751}.
%Type = Article
\bibitem[{Casas‐Ruiz et~al.(2017)Casas‐Ruiz, Catal{\'{a}}n,
  G{\'{o}}mez‐Gener, von Schiller, Obrador, Kothawala, L{\'{o}}pez, Sabater
  and Marc{\'{e}}}]{Casas-Ruiz2017}
\bibinfo{author}{Casas‐Ruiz, J.P.}, \bibinfo{author}{Catal{\'{a}}n, N.},
  \bibinfo{author}{G{\'{o}}mez‐Gener, L.}, \bibinfo{author}{von Schiller,
  D.}, \bibinfo{author}{Obrador, B.}, \bibinfo{author}{Kothawala, D.N.},
  \bibinfo{author}{L{\'{o}}pez, P.}, \bibinfo{author}{Sabater, S.},
  \bibinfo{author}{Marc{\'{e}}, R.}, \bibinfo{year}{2017}.
\newblock \bibinfo{title}{{A tale of pipes and reactors: Controls on the
  in‐stream dynamics of dissolved organic matter in rivers}}.
\newblock \bibinfo{journal}{Limnol. Oceanogr.} \bibinfo{volume}{62},
  \bibinfo{pages}{S85--S94}.
\newblock \DOIprefix\doi{10.1002/lno.10471}.
%Type = Article
\bibitem[{Coles and Jones(2000)}]{Coles2000}
\bibinfo{author}{Coles, J.F.}, \bibinfo{author}{Jones, R.C.},
  \bibinfo{year}{2000}.
\newblock \bibinfo{title}{{Effect of temperature on photosynthesis-light
  response and growth of four phytoplankton species isolated from a tidal
  freshwater river}}.
\newblock \bibinfo{journal}{J. Phycol.} \bibinfo{volume}{36},
  \bibinfo{pages}{7--16}.
\newblock \DOIprefix\doi{10.1046/j.1529-8817.2000.98219.x}.
%Type = Article
\bibitem[{Eger and Baum(2020)}]{Eger2020}
\bibinfo{author}{Eger, A.M.}, \bibinfo{author}{Baum, J.K.},
  \bibinfo{year}{2020}.
\newblock \bibinfo{title}{Trophic cascades and connectivity in coastal benthic
  marine ecosystems: a meta-analysis of experimental and observational
  research}.
\newblock \bibinfo{journal}{Marine Ecology Progress Series}
  \bibinfo{volume}{656}, \bibinfo{pages}{139--152}.
%Type = Article
\bibitem[{Elovaara et~al.(2020)Elovaara, Degerlund, Franklin, Kaartokallio and
  Tamelander}]{Elovaara2020}
\bibinfo{author}{Elovaara, S.}, \bibinfo{author}{Degerlund, M.},
  \bibinfo{author}{Franklin, D.J.}, \bibinfo{author}{Kaartokallio, H.},
  \bibinfo{author}{Tamelander, T.}, \bibinfo{year}{2020}.
\newblock \bibinfo{title}{{Seasonal variation in estuarine phytoplankton
  viability and its relationship with carbon dynamics in the Baltic Sea}}.
\newblock \bibinfo{journal}{Hydrobiologia} \bibinfo{volume}{847},
  \bibinfo{pages}{2485--2501}.
\newblock \DOIprefix\doi{10.1007/s10750-020-04267-1}.
%Type = Techreport
\bibitem[{{European Commission}(2006)}]{EC2006}
\bibinfo{author}{{European Commission}}, \bibinfo{year}{2006}.
\newblock \bibinfo{title}{{Indicators and methods for the ecological status
  assessment under the Water Framework Directive}}.
\newblock \bibinfo{type}{Technical Report}. Directorate-General Joint Research
  Centre Institute for Environment and Sustainability.
  \bibinfo{address}{Luxembourg}.
%Type = Article
\bibitem[{Fan et~al.(2025)Fan, Pein, Chen, Staneva and Cheng}]{Fan2025}
\bibinfo{author}{Fan, H.}, \bibinfo{author}{Pein, J.}, \bibinfo{author}{Chen,
  W.}, \bibinfo{author}{Staneva, J.}, \bibinfo{author}{Cheng, H.},
  \bibinfo{year}{2025}.
\newblock \bibinfo{title}{{Effects of heatwave events on dissolved oxygen in
  the Elbe Estuary}}.
\newblock \bibinfo{journal}{Water Res.} \bibinfo{volume}{286},
  \bibinfo{pages}{124125}.
\newblock \DOIprefix\doi{10.1016/j.watres.2025.124125}.
%Type = Article
\bibitem[{Ferreira et~al.(2022)Ferreira, Grigoropoulou, Saiz and
  Calbet}]{Ferreira2022}
\bibinfo{author}{Ferreira, G.D.}, \bibinfo{author}{Grigoropoulou, A.},
  \bibinfo{author}{Saiz, E.}, \bibinfo{author}{Calbet, A.},
  \bibinfo{year}{2022}.
\newblock \bibinfo{title}{{The effect of short-term temperature exposure on
  vital physiological processes of mixoplankton and protozooplankton}}.
\newblock \bibinfo{journal}{Mar. Environ. Res.} \bibinfo{volume}{179},
  \bibinfo{pages}{105693}.
\newblock \DOIprefix\doi{10.1016/j.marenvres.2022.105693}.
%Type = Misc
\bibitem[{Garc{\'{i}}a-Oliva and Lemmen(2025)}]{Garcia-Oliva2025-oxypom}
\bibinfo{author}{Garc{\'{i}}a-Oliva, O.}, \bibinfo{author}{Lemmen, C.},
  \bibinfo{year}{2025}.
\newblock \bibinfo{title}{{FABM OxyPOM and DiaMO: simple models for dissolved
  oxygen and biogeochemistry}}.
\newblock \DOIprefix\doi{10.5281/zenodo.15111434}.
%Type = Article
\bibitem[{Garnier et~al.(1999)Garnier, Billen and Palfner}]{Garnier1999}
\bibinfo{author}{Garnier, J.}, \bibinfo{author}{Billen, G.},
  \bibinfo{author}{Palfner, L.}, \bibinfo{year}{1999}.
\newblock \bibinfo{title}{{Understanding the oxygen budget and related
  ecological processes in the river Mosel: The RIVERSTRAHLER approach}}.
\newblock \bibinfo{journal}{Hydrobiologia} \bibinfo{volume}{410},
  \bibinfo{pages}{151--166}.
\newblock \DOIprefix\doi{10.1023/A:1003894200796}.
%Type = Article
\bibitem[{Graham et~al.(2024)Graham, Bierkens and van Vliet}]{Graham2024}
\bibinfo{author}{Graham, D.J.}, \bibinfo{author}{Bierkens, M.F.},
  \bibinfo{author}{van Vliet, M.T.}, \bibinfo{year}{2024}.
\newblock \bibinfo{title}{{Impacts of droughts and heatwaves on river water
  quality worldwide}}.
\newblock \bibinfo{journal}{J. Hydrol.} \bibinfo{volume}{629},
  \bibinfo{pages}{130590}.
\newblock \DOIprefix\doi{10.1016/j.jhydrol.2023.130590}.
%Type = Article
\bibitem[{Grzyb and Kulczyk(2023)}]{Grzyb2023}
\bibinfo{author}{Grzyb, T.}, \bibinfo{author}{Kulczyk, S.},
  \bibinfo{year}{2023}.
\newblock \bibinfo{title}{{How do ephemeral factors shape recreation along the
  urban river? A social media perspective}}.
\newblock \bibinfo{journal}{Landsc. Urban Plan.} \bibinfo{volume}{230},
  \bibinfo{pages}{104638}.
\newblock \DOIprefix\doi{10.1016/j.landurbplan.2022.104638}.
%Type = Article
\bibitem[{He et~al.(2024)He, Boehringer, Sch{\"{a}}fer, Heppell and
  Beck}]{He2024}
\bibinfo{author}{He, H.}, \bibinfo{author}{Boehringer, T.},
  \bibinfo{author}{Sch{\"{a}}fer, B.}, \bibinfo{author}{Heppell, K.},
  \bibinfo{author}{Beck, C.}, \bibinfo{year}{2024}.
\newblock \bibinfo{title}{{Analyzing spatio-temporal dynamics of dissolved
  oxygen for the River Thames using superstatistical methods and machine
  learning}}.
\newblock \bibinfo{journal}{Sci. Rep.} \bibinfo{volume}{14},
  \bibinfo{pages}{1--17}.
\newblock \DOIprefix\doi{10.1038/s41598-024-72084-w}.
%Type = Article
\bibitem[{Hein et~al.(2018)Hein, Viergutz, Wyrwa, Kirchesch and
  Sch{\"{o}}l}]{Hein2018}
\bibinfo{author}{Hein, B.}, \bibinfo{author}{Viergutz, C.},
  \bibinfo{author}{Wyrwa, J.}, \bibinfo{author}{Kirchesch, V.},
  \bibinfo{author}{Sch{\"{o}}l, A.}, \bibinfo{year}{2018}.
\newblock \bibinfo{title}{{Impacts of climate change on the water quality of
  the Elbe Estuary (Germany)}}.
\newblock \bibinfo{journal}{J. Appl. Water Eng. Res.} \bibinfo{volume}{6},
  \bibinfo{pages}{28--39}.
\newblock \DOIprefix\doi{10.1080/23249676.2016.1209438}.
%Type = Article
\bibitem[{Holzwarth and Wirtz(2018)}]{Holzwarth2018a}
\bibinfo{author}{Holzwarth, I.}, \bibinfo{author}{Wirtz, K.},
  \bibinfo{year}{2018}.
\newblock \bibinfo{title}{{Anthropogenic impacts on estuarine oxygen dynamics:
  A model based evaluation}}.
\newblock \bibinfo{journal}{Estuar. Coast. Shelf Sci.} \bibinfo{volume}{211},
  \bibinfo{pages}{45--61}.
\newblock \DOIprefix\doi{10.1016/j.ecss.2018.01.020}.
%Type = Article
\bibitem[{Huang et~al.(2017)Huang, Yin, Chapra and Zhou}]{Huang2017a}
\bibinfo{author}{Huang, J.}, \bibinfo{author}{Yin, H.},
  \bibinfo{author}{Chapra, S.}, \bibinfo{author}{Zhou, Q.},
  \bibinfo{year}{2017}.
\newblock \bibinfo{title}{{Modelling Dissolved Oxygen Depression in an Urban
  River in China}}.
\newblock \bibinfo{journal}{Water} \bibinfo{volume}{9}, \bibinfo{pages}{520}.
\newblock \DOIprefix\doi{10.3390/w9070520}.
%Type = Article
\bibitem[{Ingeniero et~al.(2024)Ingeniero, D{\"a}hnke, Sanders
  et~al.}]{Ingeniero2024}
\bibinfo{author}{Ingeniero, R.C.O.}, \bibinfo{author}{D{\"a}hnke, K.},
  \bibinfo{author}{Sanders, T.}, et~al., \bibinfo{year}{2024}.
\newblock \bibinfo{title}{Dissolved nitric oxide in the lower elbe estuary and
  the port of hamburg area}.
\newblock \bibinfo{journal}{Biogeosciences} \bibinfo{volume}{21},
  \bibinfo{pages}{3425--3448}.
\newblock \DOIprefix\doi{10.5194/bg-21-3425-2024}.
%Type = Article
\bibitem[{Kamjunke et~al.(2021)Kamjunke, Rode, Baborowski, Kunz, Zehner,
  Borchardt and Weitere}]{Kamjunke2021}
\bibinfo{author}{Kamjunke, N.}, \bibinfo{author}{Rode, M.},
  \bibinfo{author}{Baborowski, M.}, \bibinfo{author}{Kunz, J.V.},
  \bibinfo{author}{Zehner, J.}, \bibinfo{author}{Borchardt, D.},
  \bibinfo{author}{Weitere, M.}, \bibinfo{year}{2021}.
\newblock \bibinfo{title}{{High irradiation and low discharge promote the
  dominant role of phytoplankton in riverine nutrient dynamics}}.
\newblock \bibinfo{journal}{Limnol. Oceanogr.} \bibinfo{volume}{66},
  \bibinfo{pages}{2648--2660}.
\newblock \DOIprefix\doi{10.1002/lno.11778}.
%Type = Article
\bibitem[{Kerner(2007)}]{Kerner2007}
\bibinfo{author}{Kerner, M.}, \bibinfo{year}{2007}.
\newblock \bibinfo{title}{{Effects of deepening the Elbe Estuary on sediment
  regime and water quality}}.
\newblock \bibinfo{journal}{Estuar. Coast. Shelf Sci.} \bibinfo{volume}{75},
  \bibinfo{pages}{492--500}.
\newblock \DOIprefix\doi{10.1016/j.ecss.2007.05.033}.
%Type = Article
\bibitem[{Koll et~al.(2024)Koll, Theilen, Hauten, Woodhouse, Thiel,
  M{\"{o}}llmann and Fabrizius}]{Koll2024a}
\bibinfo{author}{Koll, R.}, \bibinfo{author}{Theilen, J.},
  \bibinfo{author}{Hauten, E.}, \bibinfo{author}{Woodhouse, J.N.},
  \bibinfo{author}{Thiel, R.}, \bibinfo{author}{M{\"{o}}llmann, C.},
  \bibinfo{author}{Fabrizius, A.}, \bibinfo{year}{2024}.
\newblock \bibinfo{title}{{Network-based integration of omics, physiological
  and environmental data in real-world Elbe estuarine Zander}}.
\newblock \bibinfo{journal}{Sci. Total Environ.} \bibinfo{volume}{942}.
\newblock \DOIprefix\doi{10.1016/j.scitotenv.2024.173656}.
%Type = Article
\bibitem[{Krause et~al.(2022)Krause, Abbott, Baranov, Bernal, Blaen, Datry,
  Drummond, Fleckenstein, Velez, Hannah, Knapp, Kurz, Lewandowski, Mart{\'{i}},
  Mendoza‐Lera, Milner, Packman, Pinay, Ward and Zarnetzke}]{Krause2022}
\bibinfo{author}{Krause, S.}, \bibinfo{author}{Abbott, B.W.},
  \bibinfo{author}{Baranov, V.}, \bibinfo{author}{Bernal, S.},
  \bibinfo{author}{Blaen, P.}, \bibinfo{author}{Datry, T.},
  \bibinfo{author}{Drummond, J.}, \bibinfo{author}{Fleckenstein, J.H.},
  \bibinfo{author}{Velez, J.G.}, \bibinfo{author}{Hannah, D.M.},
  \bibinfo{author}{Knapp, J.L.A.}, \bibinfo{author}{Kurz, M.},
  \bibinfo{author}{Lewandowski, J.}, \bibinfo{author}{Mart{\'{i}}, E.},
  \bibinfo{author}{Mendoza‐Lera, C.}, \bibinfo{author}{Milner, A.},
  \bibinfo{author}{Packman, A.}, \bibinfo{author}{Pinay, G.},
  \bibinfo{author}{Ward, A.S.}, \bibinfo{author}{Zarnetzke, J.P.},
  \bibinfo{year}{2022}.
\newblock \bibinfo{title}{{Organizational Principles of Hyporheic Exchange Flow
  and Biogeochemical Cycling in River Networks Across Scales}}.
\newblock \bibinfo{journal}{Water Resour. Res.} \bibinfo{volume}{58}.
\newblock \DOIprefix\doi{10.1029/2021WR029771}.
%Type = Article
\bibitem[{Krishna et~al.(2024)Krishna, Peterson, Listmann and
  Hinners}]{Krishna2024}
\bibinfo{author}{Krishna, S.}, \bibinfo{author}{Peterson, V.},
  \bibinfo{author}{Listmann, L.}, \bibinfo{author}{Hinners, J.},
  \bibinfo{year}{2024}.
\newblock \bibinfo{title}{{Interactive effects of viral lysis and warming in a
  coastal ocean identified from an idealized ecosystem model}}.
\newblock \bibinfo{journal}{Ecol. Modell.} \bibinfo{volume}{487},
  \bibinfo{pages}{110550}.
\newblock \DOIprefix\doi{10.1016/j.ecolmodel.2023.110550}.
%Type = Article
\bibitem[{Lespez et~al.(2025)Lespez, Germaine, Gob, Tales, Thommeret,
  de~Milleville, Archaimbault and Letourneur}]{Lespez2022}
\bibinfo{author}{Lespez, L.}, \bibinfo{author}{Germaine, M.a.},
  \bibinfo{author}{Gob, F.}, \bibinfo{author}{Tales, E.},
  \bibinfo{author}{Thommeret, N.}, \bibinfo{author}{de~Milleville, L.},
  \bibinfo{author}{Archaimbault, V.}, \bibinfo{author}{Letourneur, M.},
  \bibinfo{year}{2025}.
\newblock \bibinfo{title}{{A new tool to characterise the socio-environmental
  dimensions of urban rivers: Urban river socio-environmental index}}.
\newblock \bibinfo{journal}{Landsc. Urban Plan.} \bibinfo{volume}{261},
  \bibinfo{pages}{105388}.
\newblock \DOIprefix\doi{10.1016/j.landurbplan.2025.105388}.
%Type = Article
\bibitem[{Lewis et~al.(2024)Lewis, Lau, Jane, Rose, Be'eri‐Shlevin, Burnet,
  Clayer, Feuchtmayr, Grossart, Howard, Mariash, {Delgado Martin}, North,
  Oleksy, Pilla, Smagula, Sommaruga, Steiner, Verburg, Wain, Weyhenmeyer and
  Carey}]{Lewis2024}
\bibinfo{author}{Lewis, A.S.L.}, \bibinfo{author}{Lau, M.P.},
  \bibinfo{author}{Jane, S.F.}, \bibinfo{author}{Rose, K.C.},
  \bibinfo{author}{Be'eri‐Shlevin, Y.}, \bibinfo{author}{Burnet, S.H.},
  \bibinfo{author}{Clayer, F.}, \bibinfo{author}{Feuchtmayr, H.},
  \bibinfo{author}{Grossart, H.}, \bibinfo{author}{Howard, D.W.},
  \bibinfo{author}{Mariash, H.}, \bibinfo{author}{{Delgado Martin}, J.},
  \bibinfo{author}{North, R.L.}, \bibinfo{author}{Oleksy, I.},
  \bibinfo{author}{Pilla, R.M.}, \bibinfo{author}{Smagula, A.P.},
  \bibinfo{author}{Sommaruga, R.}, \bibinfo{author}{Steiner, S.E.},
  \bibinfo{author}{Verburg, P.}, \bibinfo{author}{Wain, D.},
  \bibinfo{author}{Weyhenmeyer, G.A.}, \bibinfo{author}{Carey, C.C.},
  \bibinfo{year}{2024}.
\newblock \bibinfo{title}{{Anoxia begets anoxia: A positive feedback to the
  deoxygenation of temperate lakes}}.
\newblock \bibinfo{journal}{Glob. Chang. Biol.} \bibinfo{volume}{30},
  \bibinfo{pages}{1--19}.
\newblock \DOIprefix\doi{10.1111/gcb.17046}.
%Type = Article
\bibitem[{Li and Wirtz(2025)}]{Li2025}
\bibinfo{author}{Li, E.}, \bibinfo{author}{Wirtz, K.}, \bibinfo{year}{2025}.
\newblock \bibinfo{title}{{Aggregation of Suspended Particles Limited by
  Preferential Settling}}.
\newblock \bibinfo{journal}{J. Geophys. Res. Ocean.} \bibinfo{volume}{130}.
\newblock \DOIprefix\doi{10.1029/2025JC022471}.
%Type = Article
\bibitem[{Lyu et~al.(2025)Lyu, Shi, Liu, Xu, Liu, Yang, Peng, Qu, Zhang, Chen,
  Zhang and Gao}]{Lyu2025}
\bibinfo{author}{Lyu, J.}, \bibinfo{author}{Shi, Y.}, \bibinfo{author}{Liu,
  T.}, \bibinfo{author}{Xu, X.}, \bibinfo{author}{Liu, S.},
  \bibinfo{author}{Yang, G.}, \bibinfo{author}{Peng, D.}, \bibinfo{author}{Qu,
  Y.}, \bibinfo{author}{Zhang, S.}, \bibinfo{author}{Chen, C.},
  \bibinfo{author}{Zhang, Y.}, \bibinfo{author}{Gao, J.}, \bibinfo{year}{2025}.
\newblock \bibinfo{title}{{Extreme drought-heatwave events threaten the
  biodiversity and stability of aquatic plankton communities in the Yangtze
  River ecosystems}}.
\newblock \bibinfo{journal}{Commun. Earth Environ.} \bibinfo{volume}{6},
  \bibinfo{pages}{1--12}.
\newblock \DOIprefix\doi{10.1038/s43247-025-02143-1}.
%Type = Article
\bibitem[{Ma et~al.(2024)Ma, Chen, Wang, Xu, Jia, Li and Hu}]{Ma2024a}
\bibinfo{author}{Ma, R.}, \bibinfo{author}{Chen, Z.}, \bibinfo{author}{Wang,
  B.}, \bibinfo{author}{Xu, C.}, \bibinfo{author}{Jia, Z.},
  \bibinfo{author}{Li, L.}, \bibinfo{author}{Hu, J.}, \bibinfo{year}{2024}.
\newblock \bibinfo{title}{{Spatiotemporal variations and controlling mechanism
  of low dissolved oxygen in a highly urbanized complex river system}}.
\newblock \bibinfo{journal}{J. Hydrol. Reg. Stud.} \bibinfo{volume}{52},
  \bibinfo{pages}{101691}.
\newblock \DOIprefix\doi{10.1016/j.ejrh.2024.101691}.
%Type = Article
\bibitem[{Mallin et~al.(2006)Mallin, Johnson, Ensign and
  MacPherson}]{Mallin2006}
\bibinfo{author}{Mallin, M.A.}, \bibinfo{author}{Johnson, V.L.},
  \bibinfo{author}{Ensign, S.H.}, \bibinfo{author}{MacPherson, T.A.},
  \bibinfo{year}{2006}.
\newblock \bibinfo{title}{{Factors contributing to hypoxia in rivers, lakes,
  and streams}}.
\newblock \bibinfo{journal}{Limnol. Oceanogr.} \bibinfo{volume}{51},
  \bibinfo{pages}{690--701}.
\newblock \DOIprefix\doi{10.4319/lo.2006.51.1_part_2.0690}.
%Type = Article
\bibitem[{Martens et~al.(2024a)Martens, Biederbick and Schaum}]{Martens2024}
\bibinfo{author}{Martens, N.}, \bibinfo{author}{Biederbick, J.},
  \bibinfo{author}{Schaum, C.E.}, \bibinfo{year}{2024}a.
\newblock \bibinfo{title}{{Picophytoplankton prevail year-round in the Elbe
  estuary}}.
\newblock \bibinfo{journal}{Plant-Environment Interact.} \bibinfo{volume}{5},
  \bibinfo{pages}{1--11}.
\newblock \DOIprefix\doi{10.1002/pei3.70014}.
%Type = Article
\bibitem[{Martens et~al.(2024b)Martens, Russnak, Woodhouse, Grossart and
  Schaum}]{Martens2024a}
\bibinfo{author}{Martens, N.}, \bibinfo{author}{Russnak, V.},
  \bibinfo{author}{Woodhouse, J.}, \bibinfo{author}{Grossart, H.P.},
  \bibinfo{author}{Schaum, C.E.}, \bibinfo{year}{2024}b.
\newblock \bibinfo{title}{{Metabarcoding reveals potentially mixotrophic
  flagellates and picophytoplankton as key groups of phytoplankton in the Elbe
  estuary}}.
\newblock \bibinfo{journal}{Environ. Res.} \bibinfo{volume}{252},
  \bibinfo{pages}{119126}.
\newblock \DOIprefix\doi{10.1016/j.envres.2024.119126}.
%Type = Article
\bibitem[{van Neer et~al.(2023)van Neer, Nachtsheim, Siebert and
  Taupp}]{VanNeer2023}
\bibinfo{author}{van Neer, A.}, \bibinfo{author}{Nachtsheim, D.},
  \bibinfo{author}{Siebert, U.}, \bibinfo{author}{Taupp, T.},
  \bibinfo{year}{2023}.
\newblock \bibinfo{title}{{Movements and spatial usage of harbour seals in the
  Elbe estuary in Germany}}.
\newblock \bibinfo{journal}{Sci. Rep.} \bibinfo{volume}{13},
  \bibinfo{pages}{1--17}.
\newblock \DOIprefix\doi{10.1038/s41598-023-33594-1}.
%Type = Article
\bibitem[{Neumann et~al.(2022)Neumann, Radtke, Cahill, Schmidt and
  Rehder}]{Neumann2022}
\bibinfo{author}{Neumann, T.}, \bibinfo{author}{Radtke, H.},
  \bibinfo{author}{Cahill, B.}, \bibinfo{author}{Schmidt, M.},
  \bibinfo{author}{Rehder, G.}, \bibinfo{year}{2022}.
\newblock \bibinfo{title}{{Non-Redfieldian carbon model for the Baltic Sea
  (ERGOM version 1.2) – implementation and budget estimates}}.
\newblock \bibinfo{journal}{Geosci. Model Dev.} \bibinfo{volume}{15},
  \bibinfo{pages}{8473--8540}.
\newblock \DOIprefix\doi{10.5194/gmd-15-8473-2022}.
%Type = Article
\bibitem[{Pace et~al.(1999)Pace, Cole, Carpenter and Kitchell}]{Pace1999}
\bibinfo{author}{Pace, M.L.}, \bibinfo{author}{Cole, J.J.},
  \bibinfo{author}{Carpenter, S.R.}, \bibinfo{author}{Kitchell, J.F.},
  \bibinfo{year}{1999}.
\newblock \bibinfo{title}{Trophic cascades revealed in diverse ecosystems}.
\newblock \bibinfo{journal}{Trends in ecology \& evolution}
  \bibinfo{volume}{14}, \bibinfo{pages}{483--488}.
%Type = Article
\bibitem[{Padhy and Singh(1977)}]{Padhy1977}
\bibinfo{author}{Padhy, R.N.}, \bibinfo{author}{Singh, P.K.},
  \bibinfo{year}{1977}.
\newblock \bibinfo{title}{{Effect of temperature on the adsorption and one-step
  growth of the Nostoc virus N-1}}.
\newblock \bibinfo{journal}{Arch. Microbiol.} \bibinfo{volume}{115},
  \bibinfo{pages}{163--167}.
\newblock \DOIprefix\doi{10.1007/BF00406370}.
%Type = Article
\bibitem[{Pein et~al.(2021)Pein, Eisele, Sanders, Daewel, Stanev, van Beusekom,
  Staneva and Schrum}]{Pein2021}
\bibinfo{author}{Pein, J.}, \bibinfo{author}{Eisele, A.},
  \bibinfo{author}{Sanders, T.}, \bibinfo{author}{Daewel, U.},
  \bibinfo{author}{Stanev, E.V.}, \bibinfo{author}{van Beusekom, J.E.},
  \bibinfo{author}{Staneva, J.}, \bibinfo{author}{Schrum, C.},
  \bibinfo{year}{2021}.
\newblock \bibinfo{title}{{Seasonal Stratification and Biogeochemical Turnover
  in the Freshwater Reach of a Partially Mixed Dredged Estuary}}.
\newblock \bibinfo{journal}{Front. Mar. Sci.} \bibinfo{volume}{8}.
\newblock \DOIprefix\doi{10.3389/fmars.2021.623714}.
%Type = Article
\bibitem[{Pein et~al.(2025)Pein, Staneva, Biederbick and Schrum}]{Pein2025}
\bibinfo{author}{Pein, J.}, \bibinfo{author}{Staneva, J.},
  \bibinfo{author}{Biederbick, J.}, \bibinfo{author}{Schrum, C.},
  \bibinfo{year}{2025}.
\newblock \bibinfo{title}{{Model-based assessment of sustainable adaptation
  options for an industrialised meso‑tidal estuary}}.
\newblock \bibinfo{journal}{Ocean Model.} \bibinfo{volume}{194},
  \bibinfo{pages}{102467}.
\newblock \DOIprefix\doi{10.1016/j.ocemod.2024.102467}.
%Type = Article
\bibitem[{Platt et~al.(1980)Platt, Gallegos and Harrison}]{Platt1980}
\bibinfo{author}{Platt, T.}, \bibinfo{author}{Gallegos, C.},
  \bibinfo{author}{Harrison, W.}, \bibinfo{year}{1980}.
\newblock \bibinfo{title}{{Photoinhibition of photosynthesis in natural
  assemblages of marine phytoplankton}}.
\newblock \bibinfo{journal}{J. Mar. Res.} \bibinfo{volume}{38},
  \bibinfo{pages}{687--701}.
%Type = Masterthesis
\bibitem[{Preußler(2025)}]{Preussler2025}
\bibinfo{author}{Preußler, N.}, \bibinfo{year}{2025}.
\newblock \bibinfo{title}{Nitrous oxide production and emission in the oxygen
  minimum zone of the Elbe estuary: A coupled hydrodynamic-biogeochemical
  model}.
\newblock \bibinfo{type}{Bachelor's thesis}. Leuphana University L{\"u}neburg.
  \bibinfo{address}{L{\"u}neburg, Germany}.
%Type = Article
\bibitem[{Raymond and Cole(2001)}]{Raymond2001}
\bibinfo{author}{Raymond, P.A.}, \bibinfo{author}{Cole, J.J.},
  \bibinfo{year}{2001}.
\newblock \bibinfo{title}{{Gas exchange in rivers and estuaries: Choosing a gas
  transfer velocity}}.
\newblock \bibinfo{journal}{Estuaries} \bibinfo{volume}{24},
  \bibinfo{pages}{312--317}.
\newblock \DOIprefix\doi{10.2307/1352954}.
%Type = Article
\bibitem[{Reese et~al.(2024)Reese, Gr{\"{a}}we, Klingbeil, Li, Lorenz and
  Burchard}]{Reese2024}
\bibinfo{author}{Reese, L.}, \bibinfo{author}{Gr{\"{a}}we, U.},
  \bibinfo{author}{Klingbeil, K.}, \bibinfo{author}{Li, X.},
  \bibinfo{author}{Lorenz, M.}, \bibinfo{author}{Burchard, H.},
  \bibinfo{year}{2024}.
\newblock \bibinfo{title}{{Local Mixing Determines Spatial Structure of
  Diahaline Exchange Flow in a Mesotidal Estuary: A Study of Extreme Runoff
  Conditions}}.
\newblock \bibinfo{journal}{J. Phys. Oceanogr.} \bibinfo{volume}{54},
  \bibinfo{pages}{3--27}.
\newblock \DOIprefix\doi{10.1175/JPO-D-23-0052.1}.
%Type = Article
\bibitem[{Rettig and Smith(2021)}]{Rettig2021}
\bibinfo{author}{Rettig, J.E.}, \bibinfo{author}{Smith, G.R.},
  \bibinfo{year}{2021}.
\newblock \bibinfo{title}{{Relative strength of top-down effects of an invasive
  fish and bottom-up effects of nutrient addition in a simple aquatic food
  web}}.
\newblock \bibinfo{journal}{Environ. Sci. Pollut. Res.} \bibinfo{volume}{28},
  \bibinfo{pages}{5845--5853}.
\newblock \DOIprefix\doi{10.1007/s11356-020-10933-7}.
%Type = Article
\bibitem[{Rewrie et~al.(2025)Rewrie, Baschek, van Beusekom, K{\"{o}}rtzinger,
  Petersen, R{\"{o}}ttgers and Voynova}]{Rewrie2025}
\bibinfo{author}{Rewrie, L.C.}, \bibinfo{author}{Baschek, B.},
  \bibinfo{author}{van Beusekom, J.E.}, \bibinfo{author}{K{\"{o}}rtzinger, A.},
  \bibinfo{author}{Petersen, W.}, \bibinfo{author}{R{\"{o}}ttgers, R.},
  \bibinfo{author}{Voynova, Y.G.}, \bibinfo{year}{2025}.
\newblock \bibinfo{title}{{Impact of primary production and net ecosystem
  metabolism on carbon and nutrient cycling at the land-sea interface}}.
\newblock \bibinfo{journal}{Front. Mar. Sci.} \bibinfo{volume}{12},
  \bibinfo{pages}{1--21}.
\newblock \DOIprefix\doi{10.3389/fmars.2025.1548463}.
%Type = Article
\bibitem[{Salk et~al.(2016)Salk, Ostrom, Biddanda, Weinke, Kendall and
  Ostrom}]{Salk2016}
\bibinfo{author}{Salk, K.R.}, \bibinfo{author}{Ostrom, P.H.},
  \bibinfo{author}{Biddanda, B.A.}, \bibinfo{author}{Weinke, A.D.},
  \bibinfo{author}{Kendall, S.T.}, \bibinfo{author}{Ostrom, N.E.},
  \bibinfo{year}{2016}.
\newblock \bibinfo{title}{{Ecosystem metabolism and greenhouse gas production
  in a mesotrophic northern temperate lake experiencing seasonal hypoxia}}.
\newblock \bibinfo{journal}{Biogeochemistry} \bibinfo{volume}{131},
  \bibinfo{pages}{303--319}.
\newblock \DOIprefix\doi{10.1007/s10533-016-0280-y}.
%Type = Article
\bibitem[{Sanders et~al.(2018)Sanders, Sch{\"{o}}l and
  D{\"{a}}hnke}]{Sanders2018}
\bibinfo{author}{Sanders, T.}, \bibinfo{author}{Sch{\"{o}}l, A.},
  \bibinfo{author}{D{\"{a}}hnke, K.}, \bibinfo{year}{2018}.
\newblock \bibinfo{title}{{Hot Spots of Nitrification in the Elbe Estuary and
  Their Impact on Nitrate Regeneration}}.
\newblock \bibinfo{journal}{Estuaries and Coasts} \bibinfo{volume}{41},
  \bibinfo{pages}{128--138}.
\newblock \DOIprefix\doi{10.1007/s12237-017-0264-8}.
%Type = Techreport
\bibitem[{Schaffrin et~al.(2021)Schaffrin, Niggemeier, Ranft, Bohne,
  Feinendegen, Sternberger, Bernhardt and Lechne}]{Schaffrin2021}
\bibinfo{author}{Schaffrin, A.}, \bibinfo{author}{Niggemeier, J.},
  \bibinfo{author}{Ranft, F.}, \bibinfo{author}{Bohne, M.},
  \bibinfo{author}{Feinendegen, L.}, \bibinfo{author}{Sternberger, J.},
  \bibinfo{author}{Bernhardt, T.}, \bibinfo{author}{Lechne, L.},
  \bibinfo{year}{2021}.
\newblock \bibinfo{title}{{Analysis of public confidence in scientific results
  related to northern European estuaries}}.
\newblock \bibinfo{type}{Technical Report}. ifok Gmbh.
  \bibinfo{address}{D{\"{u}}sseldorf}.
%Type = Article
\bibitem[{Scharfe et~al.(2009)Scharfe, Callies, Bl{\"{o}}cker, Petersen and
  Schroeder}]{Scharfe2009}
\bibinfo{author}{Scharfe, M.}, \bibinfo{author}{Callies, U.},
  \bibinfo{author}{Bl{\"{o}}cker, G.}, \bibinfo{author}{Petersen, W.},
  \bibinfo{author}{Schroeder, F.}, \bibinfo{year}{2009}.
\newblock \bibinfo{title}{{A simple Lagrangian model to simulate temporal
  variability of algae in the Elbe River}}.
\newblock \bibinfo{journal}{Ecol. Modell.} \bibinfo{volume}{220},
  \bibinfo{pages}{2173--2186}.
\newblock \DOIprefix\doi{10.1016/j.ecolmodel.2009.04.048}.
%Type = Article
\bibitem[{Sch{\"{o}}l et~al.(2014)Sch{\"{o}}l, Hein, Wyrwa, Kirchesch and
  Volker}]{Scbol2014}
\bibinfo{author}{Sch{\"{o}}l, A..}, \bibinfo{author}{Hein, B..},
  \bibinfo{author}{Wyrwa, J..}, \bibinfo{author}{Kirchesch},
  \bibinfo{author}{Volker}, \bibinfo{year}{2014}.
\newblock \bibinfo{title}{{Modelling Water Quality in the Elbe and its Estuary
  Large Scale and Long Term Applications with Focus on the Oxygen Budget of the
  Estuary}}.
\newblock \bibinfo{journal}{Kuste} , \bibinfo{pages}{203--232}.
%Type = Article
\bibitem[{Schroeder(1997)}]{Schroeder1997}
\bibinfo{author}{Schroeder, F.}, \bibinfo{year}{1997}.
\newblock \bibinfo{title}{{Water quality in the Elbe estuary: Significance of
  different processes for the oxygen deficit at Hamburg}}.
\newblock \bibinfo{journal}{Environ. Model. Assess.} \bibinfo{volume}{2},
  \bibinfo{pages}{73--82}.
\newblock \DOIprefix\doi{10.1023/a:1019032504922}.
%Type = Article
\bibitem[{Schulz et~al.(2023)Schulz, van Beusekom, Jacob, Bold, Sch{\"{o}}l,
  Ankele, Sanders and D{\"{a}}hnke}]{Schulz2023}
\bibinfo{author}{Schulz, G.}, \bibinfo{author}{van Beusekom, J.E.},
  \bibinfo{author}{Jacob, J.}, \bibinfo{author}{Bold, S.},
  \bibinfo{author}{Sch{\"{o}}l, A.}, \bibinfo{author}{Ankele, M.},
  \bibinfo{author}{Sanders, T.}, \bibinfo{author}{D{\"{a}}hnke, K.},
  \bibinfo{year}{2023}.
\newblock \bibinfo{title}{{Low discharge intensifies nitrogen retention in
  rivers – A case study in the Elbe River}}.
\newblock \bibinfo{journal}{Sci. Total Environ.} \bibinfo{volume}{904}.
\newblock \DOIprefix\doi{10.1016/j.scitotenv.2023.166740}.
%Type = Article
\bibitem[{Schwentner et~al.(2021)Schwentner, Zahiri, Yamamoto, Husemann,
  Kullmann and Thiel}]{Schwentner2021a}
\bibinfo{author}{Schwentner, M.}, \bibinfo{author}{Zahiri, R.},
  \bibinfo{author}{Yamamoto, S.}, \bibinfo{author}{Husemann, M.},
  \bibinfo{author}{Kullmann, B.}, \bibinfo{author}{Thiel, R.},
  \bibinfo{year}{2021}.
\newblock \bibinfo{title}{{eDNA as a tool for non-invasive monitoring of the
  fauna of a turbid, well-mixed system, the Elbe estuary in Germany}}.
\newblock \bibinfo{journal}{PLoS One} \bibinfo{volume}{16},
  \bibinfo{pages}{1--16}.
\newblock \DOIprefix\doi{10.1371/journal.pone.0250452}.
%Type = Article
\bibitem[{Sherman et~al.(2016)Sherman, Moore, Primeau and
  Tanouye}]{Sherman2016}
\bibinfo{author}{Sherman, E.}, \bibinfo{author}{Moore, J.K.},
  \bibinfo{author}{Primeau, F.}, \bibinfo{author}{Tanouye, D.},
  \bibinfo{year}{2016}.
\newblock \bibinfo{title}{{Temperature influence on phytoplankton community
  growth rates}}.
\newblock \bibinfo{journal}{Global Biogeochem. Cycles} \bibinfo{volume}{30},
  \bibinfo{pages}{550--559}.
\newblock \DOIprefix\doi{10.1002/2015GB005272}.
%Type = Article
\bibitem[{Spieckermann et~al.(2022)Spieckermann, Gr{\"{o}}ngr{\"{o}}ft,
  Karrasch, Neumann and Eschenbach}]{Spieckermann2022}
\bibinfo{author}{Spieckermann, M.}, \bibinfo{author}{Gr{\"{o}}ngr{\"{o}}ft,
  A.}, \bibinfo{author}{Karrasch, M.}, \bibinfo{author}{Neumann, A.},
  \bibinfo{author}{Eschenbach, A.}, \bibinfo{year}{2022}.
\newblock \bibinfo{title}{{Oxygen Consumption of Resuspended Sediments of the
  Upper Elbe Estuary: Process Identification and Prognosis}}.
\newblock \bibinfo{journal}{Aquat. Geochemistry} \bibinfo{volume}{28},
  \bibinfo{pages}{1--25}.
\newblock \DOIprefix\doi{10.1007/s10498-021-09401-6}.
%Type = Article
\bibitem[{Steidle and Vennell(2024)}]{Steidle2024}
\bibinfo{author}{Steidle, L.}, \bibinfo{author}{Vennell, R.},
  \bibinfo{year}{2024}.
\newblock \bibinfo{title}{{Phytoplankton retention mechanisms in estuaries: A
  case study of the Elbe estuary}}.
\newblock \bibinfo{journal}{Nonlinear Process. Geophys.} \bibinfo{volume}{31},
  \bibinfo{pages}{151--164}.
\newblock \DOIprefix\doi{10.5194/npg-31-151-2024}.
%Type = Article
\bibitem[{Stenström et~al.(2014)Stenström, Bakker, Brodén and
  Kvarnström}]{Stenstroem2014}
\bibinfo{author}{Stenström, F.}, \bibinfo{author}{Bakker, M.},
  \bibinfo{author}{Brodén, V.}, \bibinfo{author}{Kvarnström, E.},
  \bibinfo{year}{2014}.
\newblock \bibinfo{title}{Oxygen-induced dynamics of nitrous oxide in water and
  off-gas during the treatment of digester supernatant}.
\newblock \bibinfo{journal}{Water Science and Technology} \bibinfo{volume}{69},
  \bibinfo{pages}{84--91}.
\newblock \DOIprefix\doi{10.2166/wst.2014.673}.
%Type = Article
\bibitem[{Thongthaisong et~al.(2025)Thongthaisong, Kasada, Grossart and
  Wollrab}]{Thongthaisong2025}
\bibinfo{author}{Thongthaisong, P.}, \bibinfo{author}{Kasada, M.},
  \bibinfo{author}{Grossart, H.P.}, \bibinfo{author}{Wollrab, S.},
  \bibinfo{year}{2025}.
\newblock \bibinfo{title}{{Longer durability of host–parasite interaction
  increases host density}}.
\newblock \bibinfo{journal}{Oikos} \bibinfo{volume}{2025},
  \bibinfo{pages}{e11029}.
\newblock \DOIprefix\doi{10.1111/oik.11029}.
%Type = Article
\bibitem[{Tobias-H{\"{u}}nefeldt et~al.(2024)Tobias-H{\"{u}}nefeldt, van
  Beusekom, Russnak, D{\"{a}}hnke, Streit and Grossart}]{Tobias-hunefeldt2024}
\bibinfo{author}{Tobias-H{\"{u}}nefeldt, S.P.}, \bibinfo{author}{van Beusekom,
  J.E.}, \bibinfo{author}{Russnak, V.}, \bibinfo{author}{D{\"{a}}hnke, K.},
  \bibinfo{author}{Streit, W.R.}, \bibinfo{author}{Grossart, H.P.},
  \bibinfo{year}{2024}.
\newblock \bibinfo{title}{{Seasonality, rather than estuarine gradient or
  particle suspension/sinking dynamics, determines estuarine carbon
  distributions}}.
\newblock \bibinfo{journal}{Sci. Total Environ.} \bibinfo{volume}{926},
  \bibinfo{pages}{171962}.
\newblock \DOIprefix\doi{10.1016/j.scitotenv.2024.171962}.
%Type = Article
\bibitem[{Uber et~al.(2022)Uber, R{\"{o}}ssler, Astor, Hoffmann, {Van Oost} and
  Hillebrand}]{Uber2022}
\bibinfo{author}{Uber, M.}, \bibinfo{author}{R{\"{o}}ssler, O.},
  \bibinfo{author}{Astor, B.}, \bibinfo{author}{Hoffmann, T.},
  \bibinfo{author}{{Van Oost}, K.}, \bibinfo{author}{Hillebrand, G.},
  \bibinfo{year}{2022}.
\newblock \bibinfo{title}{{Climate Change Impacts on Soil Erosion and Sediment
  Delivery to German Federal Waterways: A Case Study of the Elbe Basin}}.
\newblock \bibinfo{journal}{Atmosphere (Basel).} \bibinfo{volume}{13},
  \bibinfo{pages}{1--21}.
\newblock \DOIprefix\doi{10.3390/atmos13111752}.
%Type = Article
\bibitem[{Verity et~al.(2002)Verity, Wassmann, Frischer, Howard-Jones and
  Allen}]{Verity2002}
\bibinfo{author}{Verity, P.G.}, \bibinfo{author}{Wassmann, P.},
  \bibinfo{author}{Frischer, M.E.}, \bibinfo{author}{Howard-Jones, M.H.},
  \bibinfo{author}{Allen, A.E.}, \bibinfo{year}{2002}.
\newblock \bibinfo{title}{{Grazing of phytoplankton by microzooplankton in the
  Barents Sea during early summer}}.
\newblock \bibinfo{journal}{J. Mar. Syst.} \bibinfo{volume}{38},
  \bibinfo{pages}{109--123}.
\newblock \DOIprefix\doi{10.1016/S0924-7963(02)00172-0}.
%Type = Article
\bibitem[{van Vliet et~al.(2023)van Vliet, Thorslund, Strokal, Hofstra,
  Fl{\"{o}}rke, {Ehalt Macedo}, Nkwasa, Tang, Kaushal, Kumar, van Griensven,
  Bouwman and Mosley}]{VanVliet2023}
\bibinfo{author}{van Vliet, M.T.}, \bibinfo{author}{Thorslund, J.},
  \bibinfo{author}{Strokal, M.}, \bibinfo{author}{Hofstra, N.},
  \bibinfo{author}{Fl{\"{o}}rke, M.}, \bibinfo{author}{{Ehalt Macedo}, H.},
  \bibinfo{author}{Nkwasa, A.}, \bibinfo{author}{Tang, T.},
  \bibinfo{author}{Kaushal, S.S.}, \bibinfo{author}{Kumar, R.},
  \bibinfo{author}{van Griensven, A.}, \bibinfo{author}{Bouwman, L.},
  \bibinfo{author}{Mosley, L.M.}, \bibinfo{year}{2023}.
\newblock \bibinfo{title}{{Global river water quality under climate change and
  hydroclimatic extremes}}.
\newblock \bibinfo{journal}{Nat. Rev. Earth Environ.} \bibinfo{volume}{4},
  \bibinfo{pages}{687--702}.
\newblock \DOIprefix\doi{10.1038/s43017-023-00472-3}.
%Type = Article
\bibitem[{Wanninkhof(1992)}]{Wanninkhof1992}
\bibinfo{author}{Wanninkhof, R.}, \bibinfo{year}{1992}.
\newblock \bibinfo{title}{{Relationship between wind speed and gas exchange
  over the ocean}}.
\newblock \bibinfo{journal}{J. Geophys. Res. Ocean.} \bibinfo{volume}{97},
  \bibinfo{pages}{7373--7382}.
\newblock \DOIprefix\doi{10.1029/92JC00188}.
%Type = Article
\bibitem[{Wanninkhof(2014)}]{Wanninkhof2014}
\bibinfo{author}{Wanninkhof, R.}, \bibinfo{year}{2014}.
\newblock \bibinfo{title}{{Relationship between wind speed and gas exchange
  over the ocean revisited}}.
\newblock \bibinfo{journal}{Limnol. Oceanogr. Methods} \bibinfo{volume}{12},
  \bibinfo{pages}{351--362}.
\newblock \DOIprefix\doi{10.4319/lom.2014.12.351}.
%Type = Article
\bibitem[{Weiss(1970)}]{Weiss1970}
\bibinfo{author}{Weiss, R.F.}, \bibinfo{year}{1970}.
\newblock \bibinfo{title}{{The solubility of nitrogen, oxygen and argon in
  water and seawater}}.
\newblock \bibinfo{journal}{Deep. Res. Oceanogr. Abstr.} \bibinfo{volume}{17},
  \bibinfo{pages}{721--735}.
\newblock \DOIprefix\doi{10.1016/0011-7471(70)90037-9}.
%Type = Article
\bibitem[{de~Wilde and de~Bie(2000)}]{deWilde2000}
\bibinfo{author}{de~Wilde, H.P.J.}, \bibinfo{author}{de~Bie, M.J.M.},
  \bibinfo{year}{2000}.
\newblock \bibinfo{title}{Nitrous oxide in the schelde estuary: production by
  nitrification and emission to the atmosphere}.
\newblock \bibinfo{journal}{Marine Chemistry} \bibinfo{volume}{69},
  \bibinfo{pages}{203--216}.
\newblock \DOIprefix\doi{10.1016/S0304-4203(99)00106-1}.
%Type = Article
\bibitem[{Wilkinson et~al.(2016)Wilkinson, Dumontier, Aalbersberg, Appleton,
  Axton, Baak, Blomberg, Boiten, da~Silva~Santos, Bourne
  et~al.}]{Wilkinson2016}
\bibinfo{author}{Wilkinson, M.D.}, \bibinfo{author}{Dumontier, M.},
  \bibinfo{author}{Aalbersberg, I.J.}, \bibinfo{author}{Appleton, G.},
  \bibinfo{author}{Axton, M.}, \bibinfo{author}{Baak, A.},
  \bibinfo{author}{Blomberg, N.}, \bibinfo{author}{Boiten, J.W.},
  \bibinfo{author}{da~Silva~Santos, L.B.}, \bibinfo{author}{Bourne, P.E.},
  et~al., \bibinfo{year}{2016}.
\newblock \bibinfo{title}{The fair guiding principles for scientific data
  management and stewardship}.
\newblock \bibinfo{journal}{Scientific data} \bibinfo{volume}{3},
  \bibinfo{pages}{1--9}.
%Type = Article
\bibitem[{Wiltshire and Manly(2004)}]{Wiltshire2004a}
\bibinfo{author}{Wiltshire, K.H.}, \bibinfo{author}{Manly, B.F.},
  \bibinfo{year}{2004}.
\newblock \bibinfo{title}{{The warming trend at Helgoland Roads, North Sea:
  Phytoplankton response}}.
\newblock \bibinfo{journal}{Helgol. Mar. Res.} \bibinfo{volume}{58},
  \bibinfo{pages}{269--273}.
\newblock \DOIprefix\doi{10.1007/s10152-004-0196-0}.
%Type = Article
\bibitem[{Wiltshire and Schroeder(1994)}]{Wiltshire1994}
\bibinfo{author}{Wiltshire, K.H.}, \bibinfo{author}{Schroeder, F.},
  \bibinfo{year}{1994}.
\newblock \bibinfo{title}{{Pigment patterns in suspended matter from the Elbe
  estuary, Northern Germany}}.
\newblock \bibinfo{journal}{Netherlands J. Aquat. Ecol.} \bibinfo{volume}{28},
  \bibinfo{pages}{255--265}.
\newblock \DOIprefix\doi{10.1007/BF02334193}.
%Type = Article
\bibitem[{Wirtz(2019)}]{Wirtz2019}
\bibinfo{author}{Wirtz, K.W.}, \bibinfo{year}{2019}.
\newblock \bibinfo{title}{{Physics or biology? Persistent chlorophyll
  accumulation in a shallow coastal sea explained by pathogens and carnivorous
  grazing}}.
\newblock \bibinfo{journal}{PLoS One} \bibinfo{volume}{14}.
\newblock \DOIprefix\doi{10.1371/journal.pone.0212143}.
%Type = Article
\bibitem[{Wirtz and Eckhardt(1996)}]{Wirtz1996}
\bibinfo{author}{Wirtz, K.W.}, \bibinfo{author}{Eckhardt, B.},
  \bibinfo{year}{1996}.
\newblock \bibinfo{title}{{Effective variables in ecosystem models with an
  application to phytoplankton succession}}.
\newblock \bibinfo{journal}{Ecol. Modell.} \bibinfo{volume}{92},
  \bibinfo{pages}{33--53}.
\newblock \DOIprefix\doi{10.1016/0304-3800(95)00196-4}.
%Type = Article
\bibitem[{Yakushev et~al.(2013)Yakushev, Debolskaya, Kuznetsov and
  Staalstr{\o}m}]{Yakushev2013}
\bibinfo{author}{Yakushev, E.V.}, \bibinfo{author}{Debolskaya, E.I.},
  \bibinfo{author}{Kuznetsov, I.S.}, \bibinfo{author}{Staalstr{\o}m, A.},
  \bibinfo{year}{2013}.
\newblock \bibinfo{title}{{Modelling of the meromictic fjord hunnbunn (Norway)
  with an oxygen depletion model (OxyDep)}}.
\newblock \bibinfo{journal}{Handb. Environ. Chem.} \bibinfo{volume}{22},
  \bibinfo{pages}{235--251}.
\newblock \DOIprefix\doi{10.1007/698{\textunderscore}2011{\textunderscore}110}.
%Type = Article
\bibitem[{Yakushev et~al.(2017)Yakushev, Protsenko, Bruggeman, Wallhead,
  Pakhomova, Yakubov, Bellerby and Couture}]{Yakushev2017}
\bibinfo{author}{Yakushev, E.V.}, \bibinfo{author}{Protsenko, E.A.},
  \bibinfo{author}{Bruggeman, J.}, \bibinfo{author}{Wallhead, P.},
  \bibinfo{author}{Pakhomova, S.V.}, \bibinfo{author}{Yakubov, S.K.},
  \bibinfo{author}{Bellerby, R.G.}, \bibinfo{author}{Couture, R.M.},
  \bibinfo{year}{2017}.
\newblock \bibinfo{title}{{Bottom RedOx Model (BROM v.1.1): A coupled
  benthic-pelagic model for simulation of water and sediment biogeochemistry}}.
\newblock \bibinfo{journal}{Geosci. Model Dev.} \bibinfo{volume}{10},
  \bibinfo{pages}{453--482}.
\newblock \DOIprefix\doi{10.5194/gmd-10-453-2017}.
%Type = Article
\bibitem[{Zeng et~al.(2019)Zeng, Chen, Guo, Lin, Mu, Fan, Zheng and
  Qiu}]{Zeng2019b}
\bibinfo{author}{Zeng, J.}, \bibinfo{author}{Chen, M.}, \bibinfo{author}{Guo,
  L.}, \bibinfo{author}{Lin, H.}, \bibinfo{author}{Mu, X.},
  \bibinfo{author}{Fan, L.}, \bibinfo{author}{Zheng, M.}, \bibinfo{author}{Qiu,
  Y.}, \bibinfo{year}{2019}.
\newblock \bibinfo{title}{{Role of organic components in regulating
  denitrification in the coastal water of Daya Bay, southern China}}.
\newblock \bibinfo{journal}{Environ. Sci. Process. Impacts}
  \bibinfo{volume}{21}, \bibinfo{pages}{831--844}.
\newblock \DOIprefix\doi{10.1039/c8em00558c}.
%Type = Article
\bibitem[{Zhi et~al.(2023)Zhi, Klingler, Liu and Li}]{Zhi2023}
\bibinfo{author}{Zhi, W.}, \bibinfo{author}{Klingler, C.},
  \bibinfo{author}{Liu, J.}, \bibinfo{author}{Li, L.}, \bibinfo{year}{2023}.
\newblock \bibinfo{title}{{Widespread deoxygenation in warming rivers}}.
\newblock \bibinfo{journal}{Nat. Clim. Chang.} \bibinfo{volume}{13},
  \bibinfo{pages}{1105--1113}.
\newblock \DOIprefix\doi{10.1038/s41558-023-01793-3}.

\end{thebibliography}
%\bibliography{/home/og/library_it}


\onecolumn
\appendix
\numberwithin{equation}{section} 
%%%%%%%%%%%%%%%%%%%%%%%%%%%%%%%%%%%%%%%%%%%%%%%%%%%%%%%%%%%%%%%%%%%%%%%%%%%%%%%
\section{Full model description}\label{sm:model-description}
%%%%%%%%%%%%%%%%%%%%%%%%%%%%%%%%%%%%%%%%%%%%%%%%%%%%%%%%%%%%%%%%%%%%%%%%%%%%%%%
\subsection{Re-aeration and water-air oxygen exchange}
Oxygen transport due to aeration is calculated using the phenomenological equations of \citet{Wanninkhof1992,Wanninkhof2014}.
The oxygen surface flux is

\begin{equation}
 \textrm{Re-aeration} = \alpha \cdot k_{\textrm{\scriptsize air}}\cdot(\textrm{Sat} - \DO),
 \label{eq:Aer}
\end{equation}

where $\alpha$ is a form factor that corrects surface to volume ratio of idealized topologies \citep{Holzwarth2018a}, oxygen is the dissolved oxygen in water, and Sat is the saturation oxygen concentration in water as function of temperature and salinity after \citep{Weiss1970}.
 $k_{\textrm{\scriptsize air}}$ is the coefficient of re-aeration, which we defined as a function by parts corrected in low wind speeds after \citep{Raymond2001}

\begin{equation}
 k_{\textrm{\scriptsize air}} = \left(\frac{\textrm{Sc}}{660}\right)^{-0.5} \cdot
 \begin{cases}
 0.0283\cdot w^3, & \textrm{if } w \geq 11\,  \textrm{m s\textsuperscript{-1}}  \\
 0.3120\cdot w^2, & \textrm{if } 11 > w \geq 3\, \textrm{m s\textsuperscript{-1}}\\
 0.9836\cdot \textrm{e}^{0.35\cdot w}, & \textrm{if } w < 3\, \textrm{m s\textsuperscript{-1}}
 \end{cases}
 \label{eq:krear}
\end{equation}

where Sc is the Schmidt number, which depends on temperature and salinity \citep{Wanninkhof2014}, and $w$ is the wind velocity in m s\textsuperscript{-1} .
This formulation is new to the model as the original implementation was a vertical-averaged description \citep{Holzwarth2018a} and here we redefined the process to be vertically explicit.

%%%%%%%%%%%%%%%%%%%%%%%%%%%%%%%%%%%%%%%%%%%%%%%%%%%%%%%%%%%%%%%%%%%%%%%%%%%%%%%
\subsection{Micro-algae}
The two micro-algae classes (one with dependence on dissolved silicate, thus representing diatoms) have growth rates that depend on photosynthesis, which is limited by light availability, dissolved nitrogen and ortho-phosphate concentrations.
 
\begin{equation}
 \frac{\textrm{d} \textrm{Alg}_i}{\textrm{d}t} =  \textrm{Alg}_i \cdot \left( \rho_i - m_i - r_i - \lambda_i \right) ,
 \label{eq:Alg}
\end{equation}

where $\rho_i$ is the gross production rate, $m_i$ is loss rate (mortality), $r_i$ is the respiration rate and $\lambda_i$ is the virus related mortality for each micro-algae.

The gross primary production is defined as

\begin{equation}
 \rho_i = f_{\textrm{\scriptsize light},i}\cdot f_{\textrm{\scriptsize nut},i}\cdot\tau_{\textrm{\scriptsize Alg},i}(T)\cdot \mu^*_i,
 \label{eq:rho}
\end{equation}

where $f_{\textrm{\scriptsize light},i}$ and $f_{\textrm{\scriptsize nut},i}$ are growth limitation coefficients for nutrient and light, $\mu^*_i$ is the maximum growth rate in optimal conditions---neither light nor nutrient limitation---at a reference temperature of $T' = 20$ $^\circ$C and $\tau_{\textrm{\scriptsize Alg},i}(T)$ is a temperature-sensitivity function defined as a $Q_{10}$ rule 

\begin{equation}
 \tau_{\textrm{\scriptsize Alg},i}(T) = Q_{10,\textrm{\scriptsize Alg},i}^{\frac{T-T^*}{10}}. 
 \label{eq:Q10}
\end{equation}

The limitation coefficients for light are defined as a saturating exponential 

\begin{equation}
 f_{\textrm{\scriptsize light},i} = 1 - \textrm{e}^{-I/I^*_i}, 
 \label{eq:flight}
\end{equation}

where $I$ is the light intensity and $I^*_i$ is a reference light intensity.
The limitation coefficients for nutrients are defined as a Leiblig's rule of the minimum

\begin{equation}
 f_{\textrm{\scriptsize nut},i} = \min\left(f_{N,i},f_{P,i},f_{Si,i}\right),
 \label{eq:fnut}
\end{equation}

where each $f_X$ follows a Monod equation with half saturations $K_X$, thus

\begin{equation}
 f_{X,i} = \frac{X}{X+K_X}.
 \label{eq:fnutX}
\end{equation}

As by convention Alg$_1$ represents diatoms---silicate limited--- and Alg$_2$ other non-diatoms, we set $f_{Si,2} = 1$ to avoid silica limitation in non-diatoms. 

The mortality rate $m_i$ is a piece-wise defined function defined after \citep{Scharfe2009}, which emulate temperature dependent grazing with inhibition during colder temperatures

\begin{equation}
 m_i = m^*_i \cdot
 \begin{cases}
 \tau_m(T), & \textrm{for } T \geq 20\,^\circ\textrm{C} \\
 1, & \textrm{for } 5 < T < 20\,^\circ\textrm{C} \\
 0.33, & \textrm{for } T \leq 5\,^\circ\textrm{C}.
 \end{cases}
 \label{eq:m}
\end{equation}

The respiration rate $r_i$ is divided in basal and maintenance respiration rates, weighted by a factor $\pi$ as

\begin{equation}
 r_i = \pi\cdot\rho + (1-\pi)\cdot r^*_i\cdot\tau_{\textrm{\scriptsize Res}}(T),
 \label{eq:r}
\end{equation}

where $r^*_i$ is the respiration rate at the reference temperature and $\tau_{\textrm{\scriptsize Res}}$ is the temperature sensitivity.

The virus related mortality or cell lysis $\lambda_i$ depends on a step-function \citep{Wirtz2019} defined as

\begin{equation}
 \lambda_i = \frac{1}{1+\textrm{e}^{S\cdot\left(1-\textrm{\footnotesize Vir}_i\right)}},
 \label{eq:sigma}
\end{equation}

where $\textrm{Vir}_i$ is the extent of the viral infection, and $S$ is a sensitivity of mortality to viral infection.

The oxygen produced by photosynthesis used in the oxygen budget Eq.~\ref{eq:DO} is the addition of the gross primary production of both micro-algae classes

\begin{equation}
 \textrm{Photosynthesis} = \sum_{i \in (1,2)} \rho_i\cdot\textrm{Alg}_i ,
 \label{eq:Pho}
\end{equation}

and the respiration rate is the sum of the respiration rates plus a rather small fraction $c$ of the mortality rate to emulate grazers respiration

\begin{equation}
 \textrm{Respiration} = \sum_{i \in (1,2)} (r_i + c\cdot m_i )\cdot\textrm{Alg}_i.
 \label{eq:Res}
\end{equation}

%%%%%%%%%%%%%%%%%%%%%%%%%%%%%%%%%%%%%%%%%%%%%%%%%%%%%%%%%%%%%%%%%%%%%%%%%%%%%%%
\subsection{Nutrient uptake}
Nutrient uptake follows a fixed stoichiometry approach where the cell composition remains constant.
The uptake rate is thus proportional to the net primary production, i.e., gross primary production minus respiration.
In the case of phosphorus and silica, which are represented as a single inorganic form, the uptake rates are straightforwardly calculated as

\begin{equation}
 u_X = a_X\cdot\sum_{i \in (1,2)} (\rho_i-r_i)\cdot\textrm{Alg}_i, 
 \label{eq:u1}
\end{equation}

$a_X$ is the stoichiometric ratio of $X$ relative to carbon.

For nitrogen, the ammonia uptake is preferred in a flexible way unless the ammonia concentration is too low when nitrate uptake are instead preferred.
The fraction of ammonia uptake when ammonia concentration is below a critical value of 0.7 mmol m\textsuperscript{-3} is

\begin{equation}
 f_{\textrm{\scriptsize NH}_4} = \frac{\textrm{\small NH}_4}{\textrm{\small NH}_4 + \textrm{\small NO}_3}, 
 \label{eq:fNH4}
\end{equation}

Ammonia and nitrate uptake is thus defined as 

\begin{eqnarray}
 u_{ \textrm{NH}_4} &=& a_N\cdot f_{\textrm{\scriptsize NH}_4}\cdot\sum_{i \in (1,2)} (\rho_i-r_i)\cdot\textrm{Alg}_i, \label{eq:uNH4} \\
 u_{\textrm{NO}_3} &=& a_N\cdot (1-f_{\textrm{\scriptsize NH}_4})\cdot\sum_{i \in (1,2)} (\rho_i-r_i)\cdot\textrm{Alg}_i.
 \label{eq:uNO3}
\end{eqnarray}

In the case of ammonia above the critical value, we assume that nitrate uptake is inhibited, thus setting $f_{\textrm{\scriptsize NH}_4} = 1$.

%%%%%%%%%%%%%%%%%%%%%%%%%%%%%%%%%%%%%%%%%%%%%%%%%%%%%%%%%%%%%%%%%%%%%%%%%%%%%%%
\subsection{Particulate and dissolved organic matter}
POM and DOM have an explicit elemental composition (carbon, nitrogen and phosphorus).
POM is present in two qualities, which transition in the sequences labile $\rightarrow$ semi-labile $\rightarrow$ dissolved and labile $\rightarrow$ dissolved.
Labile and semi-labile POM follow the dynamics:

\begin{eqnarray}
 \frac{\textrm{d} \textrm{PO}X_L}{\textrm{d}t} &=& a_X\cdot(1 - f)\cdot(m_1 + m_2 +\lambda_1 +\lambda_2) - d_{X,L\rightarrow S} - d_{X,L\rightarrow D} - M_{X,L},\label{eq:POXL}\\
 \frac{\textrm{d} \textrm{PO}X_S}{\textrm{d}t} &=& d_{X,L\rightarrow S } - d_{C,S\rightarrow D} - M_{X,S},\label{eq:POXS}
\end{eqnarray}

where $X$ is either carbon, nitrogen or phosphorus, the subindexes $L$, $S$, and $D$ are for labile, semi-labile, and dissolved, respectively, $f$ is the fraction of nutrient released by autolysis, the sum $m_1 + m_2 +\lambda_1 +\lambda_2$ is the total loss rate including mortality and lysis rates for each micro-algae, $d_{X,i\rightarrow j}$ are degradation rates from the quality $i$ to the quality $j$, and $M_{X,i}$ are mineralization rates.

POM and DOM mineralize to DOC, nitrogen and ortho-phosphate. 
Unlike for dissolved nutrients, dissolved carbon is calculated using a simple mass balance

\begin{equation}
 \frac{\textrm{d} \textrm{DOC}}{\textrm{d}t} = d_{C,L\rightarrow D} + d_{C,S\rightarrow D} - M_{C,D}.
 \label{eq:DOC}
\end{equation}

POM mineralization rates for the element $X$ of quality $i$ are defined as 
\begin{equation}
 M_{X,i} = k_{\textrm{\scriptsize Min},i}\cdot \tau_{\textrm{\scriptsize Min}}(T) \cdot \textrm{PO}X_i,
 \label{eq:MX}
\end{equation}

where $k_{\textrm{\scriptsize Min},i}$ are mineralization rate constants for the quality $i$ and $\tau_{\textrm{\scriptsize Min}}(T)$ is the temperature sensitivity.

Degradation rates are defined from the mineralization rates as 
\begin{equation}
 d_{X,i\rightarrow j} = \kappa_{i\rightarrow j} \cdot M_{X,i},
 \label{eq:dX}
\end{equation}

where $\kappa_{i\rightarrow j}$ is a factor for the decomposition from the quality $i$ to the quality $j$.

%%%%%%%%%%%%%%%%%%%%%%%%%%%%%%%%%%%%%%%%%%%%%%%%%%%%%%%%%%%%%%%%%%%%%%%%%%%%%%%
\subsection{Dissolved nitrogen, nitrification and mineral denitrification}

Dissolved inorganic nitrogen is the sum of ammonium and nitrate and ammonium transitions to nitrate as a function of oxygen. 
The mass balance for dissolved inorganic nitrogen is thus

\begin{eqnarray}
 \frac{\textrm{d} \textrm{NH}_4}{\textrm{d}t} &=& a_N\cdot f\cdot(m_1 + m_2 + \lambda_1 + \lambda_2) + M_{N} - u_{\textrm{\scriptsize NH}_4}- \gamma, \label{eq:NH4} \\
 \frac{\textrm{d} \textrm{NO}_3}{\textrm{d}t} &=& \gamma - M_{\textrm{\scriptsize NO}_3} - u_{\textrm{\scriptsize NO}_3},
 \label{eq:NO3}
\end{eqnarray}

where $\gamma$ is the ammonia consumed in the nitrification process, $M_{N}$ is the total nitrogen mineralized from POM and DOM ($= \sum_{i \in (L,S,D)} M_{N,i}$), $M_{\textrm{\scriptsize NO}_3}$ is mineral denitrification into N\textsubscript{2}, $u_{\textrm{\scriptsize NH}_4}$ and $u_{\textrm{\scriptsize NO}_3}$ are total micro-algae uptake rates.

The ammonia used during nitrification depends on ammonia and oxygen concentrations as 

\begin{equation}
 \gamma = k_{\textrm{\scriptsize Nit}}\cdot \tau_{\textrm{\scriptsize Nit}}(T)\cdot \frac{\textrm{\small NH}_4}{\textrm{\small NH}_4 + K_{\textrm{\scriptsize NH}_4}}\cdot \frac{\textrm{\small\DO}}{\textrm{\small\DO} + K^*_{\textrm{\scriptsize\DO}}} ,
 \label{eq:gamma}
\end{equation}

where $K_{\textrm{\scriptsize NH4}_4}$ and $K^*_{\textrm{\scriptsize\DO}}$ are half saturation constants for ammonia and oxygen in nitrification, $k_{\textrm{\scriptsize Nit}}$ is the nitrification rate at the reference temperature and $\tau_{\textrm{\scriptsize Nit}}(T)$ is the temperature sensitivity.

As the nitrification process consumes two moles of oxygen per each mol of nitrified ammonia in a two-step process

\begin{eqnarray*}
 \mathrm{NH_4^+ + 1.5\,O_2} &\rightarrow& \mathrm{NO_2^- + H_2O + 2H^+}\\
 \mathrm{NO_2^- + 0.5\,O_2} &\rightarrow& \mathrm{NO_3^-},
\end{eqnarray*}

the total oxygen consumed during mineralization used in the oxygen budget (Eq.~\ref{eq:DO}) is

\begin{equation}
 \textrm{Nitrification} = 2\cdot\gamma.
 \label{eq:Nit}
\end{equation}

Mineral denitrification is a fraction of total mineralized organics that is achieved by nitrate denitrification $\Phi_N$ in contrast to oxygen consumption as a competing mechanism---which is proportional to ($1-\Phi_N$)--- thus

\begin{equation}
 M_{\textrm{\scriptsize NO}_3} = \Phi_N \cdot\sum_{i \in (L,S,D)} M_{C,i} \,\, \textrm{with}\,\, \Phi_N = \frac{\phi_N}{\phi_N+\phi_{\textrm{\scriptsize\DO}}}
 \label{eq:MNO3}
\end{equation}

where $\phi_N$ and $\phi_{\textrm{\scriptsize\DO}}$ are contributions of nitrate and oxygen in mineralization defined as

\begin{eqnarray}
 \phi_N &=& \frac{\textrm{\small NO}_3}{\textrm{\small NO}_3 + K_{\textrm{\scriptsize NO}_3}}\cdot \left(1-\frac{\textrm{\small\DO}}{\textrm{\small\DO} + K_{\textrm{\scriptsize\DO}}} \right)\cdot\tau_N(T)
 \label{eq:phiN}\\
 \phi_{\textrm{\scriptsize\DO}} &=& \frac{\textrm{\small\DO}}{\textrm{\small\DO} + K'_{\textrm{\scriptsize\DO}}}\cdot\tau_{\textrm{\scriptsize\DO}}(T), 
 \label{eq:phiDO}
\end{eqnarray}

where $K_{\textrm{\scriptsize NO}_3}$ $K_{\textrm{\scriptsize\DO}}$ and $K'_{\textrm{\scriptsize\DO}}$ are half saturation constants for nitrate in denitrification, oxygen inhibition in denitrification and oxygen consumption in mineralization, and $\tau_i$ are temperature-sensitivities with an appropriate $Q_{10}$ value.

The oxygen consumed during mineralization used in the oxygen budget (Eq.~\ref{eq:DO}) is

\begin{equation}
 \textrm{Mineralization} = (1 - \Phi_N) \cdot\sum_{i \in (L,S,D)} M_{C,i}.
 \label{eq:Min}
\end{equation}

%%%%%%%%%%%%%%%%%%%%%%%%%%%%%%%%%%%%%%%%%%%%%%%%%%%%%%%%%%%%%%%%%%%%%%%%%%%%%%%
\subsection{Ortho-phosphate and silicate}
Ortho-phosphate is similar

\begin{equation}
 \frac{\textrm{d} \textrm{PO}_4}{\textrm{d}t} = a_P\cdot f\cdot(m_1 + m_2+ \lambda_1 + \lambda_2) + M_{P} - u_{\textrm{\scriptsize PO}_4}
 \label{eq:PO4}
\end{equation}

where $M_{P}$ is the total phosphorus mineralized from POM and DOM ($= \sum_{i \in (L,S,D)} M_{P,i}$), and $u_{\textrm{\scriptsize PO4}_4}$ is the total micro-algae uptake rates for ortho-phosphate.

Unlike the other nitrogen and phosphorus, silicate is present in dissolved --bio-available-- and particulate mineral (Opal) forms

\begin{eqnarray}
 \frac{\textrm{d} \textrm{\scriptsize Si}} {\textrm{d}t} &=& a_{\textrm{\scriptsize Si}}\cdot f\cdot (m_1+\lambda_1) + D_{\textrm{\scriptsize Si}} - u_{\textrm{\scriptsize Si}} \label{eq:Si},\\
 \frac{\textrm{d} \textrm{Opal}} {\textrm{d}t} &=& a_{\textrm{\scriptsize Si}}\cdot (1-f)\cdot m_1 - D_{\textrm{\scriptsize Si}} 
\end{eqnarray}

where $D_{Si}$ is the dissolution rate of opal to bio-available dissolved silicate defined as

\begin{equation}
 D_{Si} = k_{Si}\cdot\textrm{Opal}\cdot(\textrm{Si}'-\textrm{Si}),
 \label{eq:DSi}
\end{equation}

where $k_{Si}$ is opal dissolution reaction rate constant, and $\textrm{Si}'$ is a reference (saturation) bio-available silicate concentration.
Note that the equations for silicate balance include only one micro-algae class, which represents diatoms.

%%%%%%%%%%%%%%%%%%%%%%%%%%%%%%%%%%%%%%%%%%%%%%%%%%%%%%%%%%%%%%%%%%%%%%%%%%%%%%%
\subsection{Viral infection}
We consider two different virus, each one exclusively infecting either Alg$_1$ or Alg$_2$. 
The dynamics in intracellular viral density Vir$_i$ is here described by 

\begin{equation}
 \frac{\textrm{d} \textrm{Vir}_i}{\textrm{d}t} = G_i\cdot n_i - H_i - B_i,
 \label{eq:Vir}
\end{equation}

where $G_i$ is infection–replication, $n_i$ is the burst size (relative number of viral particles produced per host), $H_i$ virus removal by host mortality (lysis), and $B_i$ is virus inactivation.

The infection–replication is expressed as the probability of an infected host contacting a susceptible one. 
This is hindered by high concentrations immune and inert particles. 
Based on collision dynamics, we use 

\begin{equation}
 G_i= G^*\cdot \frac{\textrm{\small Vir}_i^2}{\textrm{\small POC}_L+\textrm{\small POC}_S+\textrm{\small Alg}_j},
 \label{eq:G}
\end{equation}

where $G^*$ is a reference infection rate.
The burst size is a temperature-dependent step-function as

\begin{equation}
 n_i= n^*\cdot \tau_n(T)\cdot\frac{1}{1+\textrm{e}^{S\cdot\left(\textrm{\footnotesize Vir}^*-\textrm{\footnotesize Vir}_i\right)}},
 \label{eq:n}
\end{equation}

where $n^*$ is the reference burst size when Vir$_i$ is large enough, $\tau_n$ is the temperature sensitivity, and Vir$^*_i$ is a half-saturation value.

Virus removal by host mortality is defined as the preferential loss of heavily infected hosts. 
This is expressed as an increased host mortality because infected phytoplankton frequently undergo apoptosis. 
This selective removal of infected individuals leads to a relative increase in the survival of healthy or less infected hosts or species, thereby reducing the average viral density.
This is expressed using trait dynamics formulation \citep{Wirtz1996}

\begin{equation}
 H_i= \textrm{Vir}_i \cdot (\textrm{Vir}^* -\textrm{Vir}_i)\cdot \textrm{e}^{-\frac{\textrm{\tiny Alg}_i}{C} }\cdot\frac{\textrm{\small Vir}_i}{\textrm{\small Vir}_i+\textrm{\small Vir}'} \cdot\lambda_i^2\cdot S\cdot \textrm{e}^{S\cdot(1-\textrm{\footnotesize Vir}_i)},
 \label{eq:H}
\end{equation}

where Vir$'$ is a reference low-value for virus abundance and $C$ is a constant that modulates plankton diversity \citep{Wirtz2019}.
The first two terms $\textrm{Vir}_i \cdot (\textrm{Vir}^*_i -\textrm{Vir}_i)$ are diversity of the infection levels as calculated as in a binomial trait \citep{Wirtz1996}, and the exponential term $\textrm{e}^{-\frac{\textrm{\tiny Alg}_i}{C} }$ emulates the genetic diversity---coupled to virus susceptibility---reduction during blooming phases.
The virus removal due to preferential loss thus depends on the diversity of the levels of infection host and of the host defense.
The term $\frac{\textrm{\small Vir}_i}{\textrm{\small Vir}_i+\textrm{\small Vir}'}$ mantains the levels of infection to viable levels \citep{Wirtz2019}.
The remaining terms are the first partial derivative of $\lambda_i$ in respect to $\textrm{Vir}_i$, standard method for trait dynamics $\frac{\partial \lambda_i}{\partial \textrm{Vir}_i} = \lambda_i^2\cdot S\cdot \textrm{e}^{S\cdot(1-\textrm{\footnotesize Vir}_i)}$

Virus inactivation $B_i$ is expressed by

\begin{equation}
 B_i= B^*\cdot \tau_B(T)\cdot\frac{\textrm{\small Vir}_i^2}{\textrm{\small Vir}_i+\textrm{\small Vir}'},
 \label{eq:B}
\end{equation}

where $B^*$ is a reference inactivation value, and $\tau_B$ is the temperature dependence.

%%%%%%%%%%%%%%%%%%%%%%%%%%%%%%%%%%%%%%%%%%%%%%%%%%%%%%%%%%%%%%%%%%%%%%%%%%%%%%%
\subsection{Light attenuation}
Light attenuation follows an exponential decay with an exponential coefficient $\zeta$ defined by

\begin{equation}
  \zeta = \zeta_0 + \epsilon_\textrm{POC}\cdot(\textrm{POC}_L+\textrm{POC}_S) + \epsilon_\textrm{Alg}\cdot(\textrm{Alg}_1+\textrm{Alg}_2) + \epsilon_\textrm{ISPM}\cdot \textrm{ISPM},  
  \label{eq:zeta}
\end{equation}

where $\zeta_0$ is the background attenuation coefficient, $\epsilon_\textrm{POC}$, $\epsilon_\textrm{Alg}$ and $\epsilon_\textrm{ISPM}$ are specific attenuation coefficients for POM, micro-algae, and a constant concentration of inorganic suspended particulate matter (ISPM), respectively.

%%%%%%%%%%%%%%%%%%%%%%%%%%%%%%%%%%%%%%%%%%%%%%%%%%%%%%%%%%%%%%%%%%%%%%%%%%%%%%%
\subsection{Oxygen flux}
The oxygen flux $f$ is defined as the vertically integrated Oxy dynamics (Eq.~\ref{eq:DO}) from the free surface ($\eta=0$) to the maximum depth $\eta_{\max}$ for each position along the river

\begin{equation}
    f = \int_{\eta=0}^{\eta_{\max}}\frac{\textrm{d} \textrm{\DO}}{\textrm{d}t} \cdot \textrm{d}z.
    \label{eq:flux}
\end{equation}

%%%%%%%%%%%%%%%%%%%%%%%%%%%%%%%%%%%%%%%%%%%%%%%%%%%%%%%%%%%%%%%%%%%%%%%%%%%%%%%
\section{Extended model validation}\label{sec:sm-validation}
\begin{figure}[htb] %htb
 \centering
 \includegraphics[page=1,width=1.\textwidth]{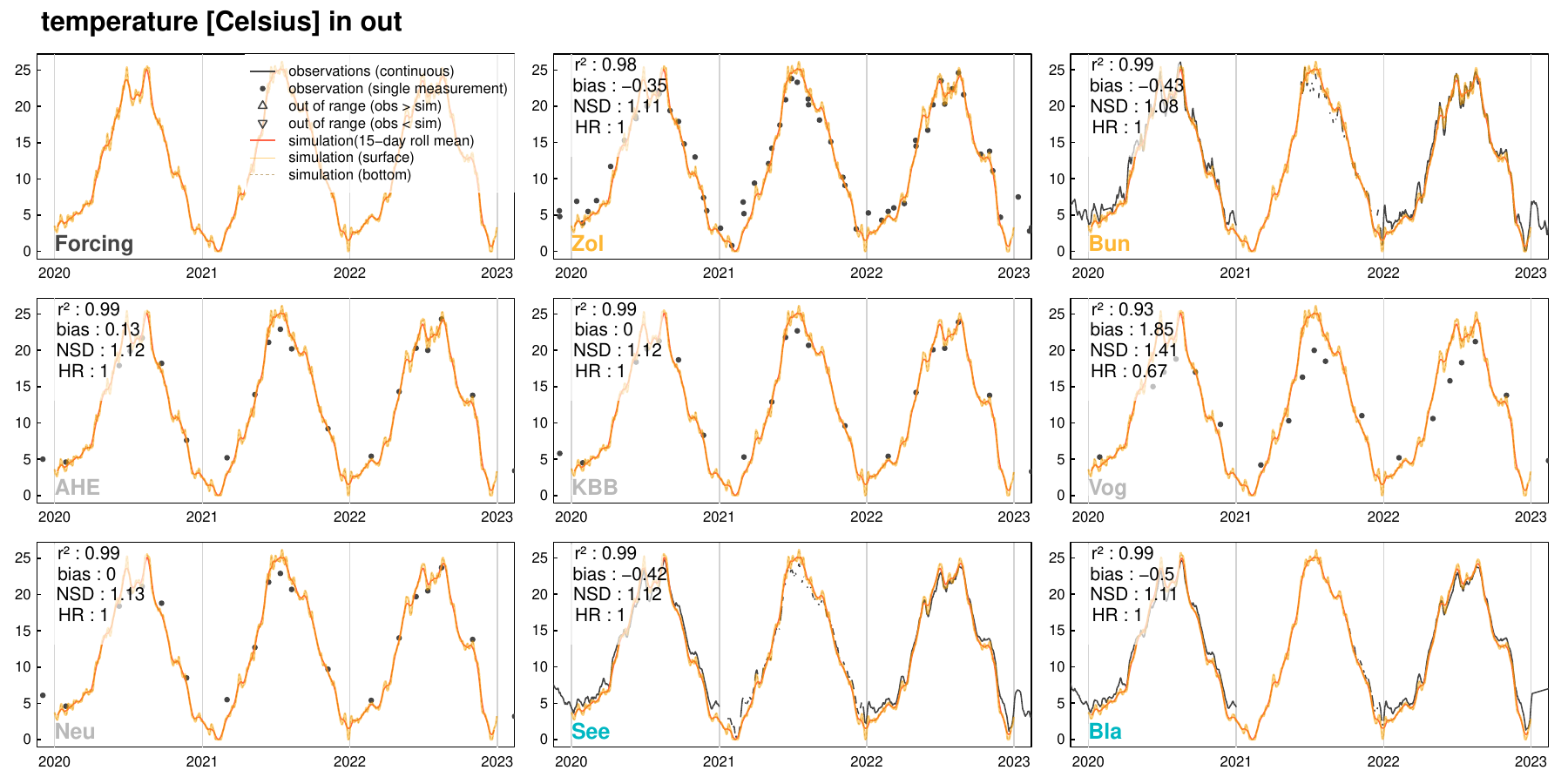} 
 \includegraphics[page=2,width=1.\textwidth]{sm-validation/out.temp.validation} 
 \caption{\textbf{Temperature validation:} 
 Comparison of modelled (orange lines) and observed (black points and lines) values used as forcing in Geesthacht Wier and in each station (Zol = Zollenspieker, Bun = Bunthaus, AHE = Alte Harburger Elbbrücke, KBB = Köhlbrandbrücke, VNE = Vogelsander Norderelbe, Neu = Neumühlen, See = Seemannshöft, and Bla = Blankenese). 
 In each plot, we included the performance metrics for each station ($r^2$ = coefficient of determination, bias, NSD = normalized standard deviation, and HR = hit rate).
 The Taylor diagram (bottom left) and the aggregated observations vs. model comparisona and their performance metrics (bottom right) are also shown.}
 \label{fig:validation-temp}
\end{figure}

\begin{figure}[htb] %htb
 \centering
 \includegraphics[page=1,width=1.\textwidth]{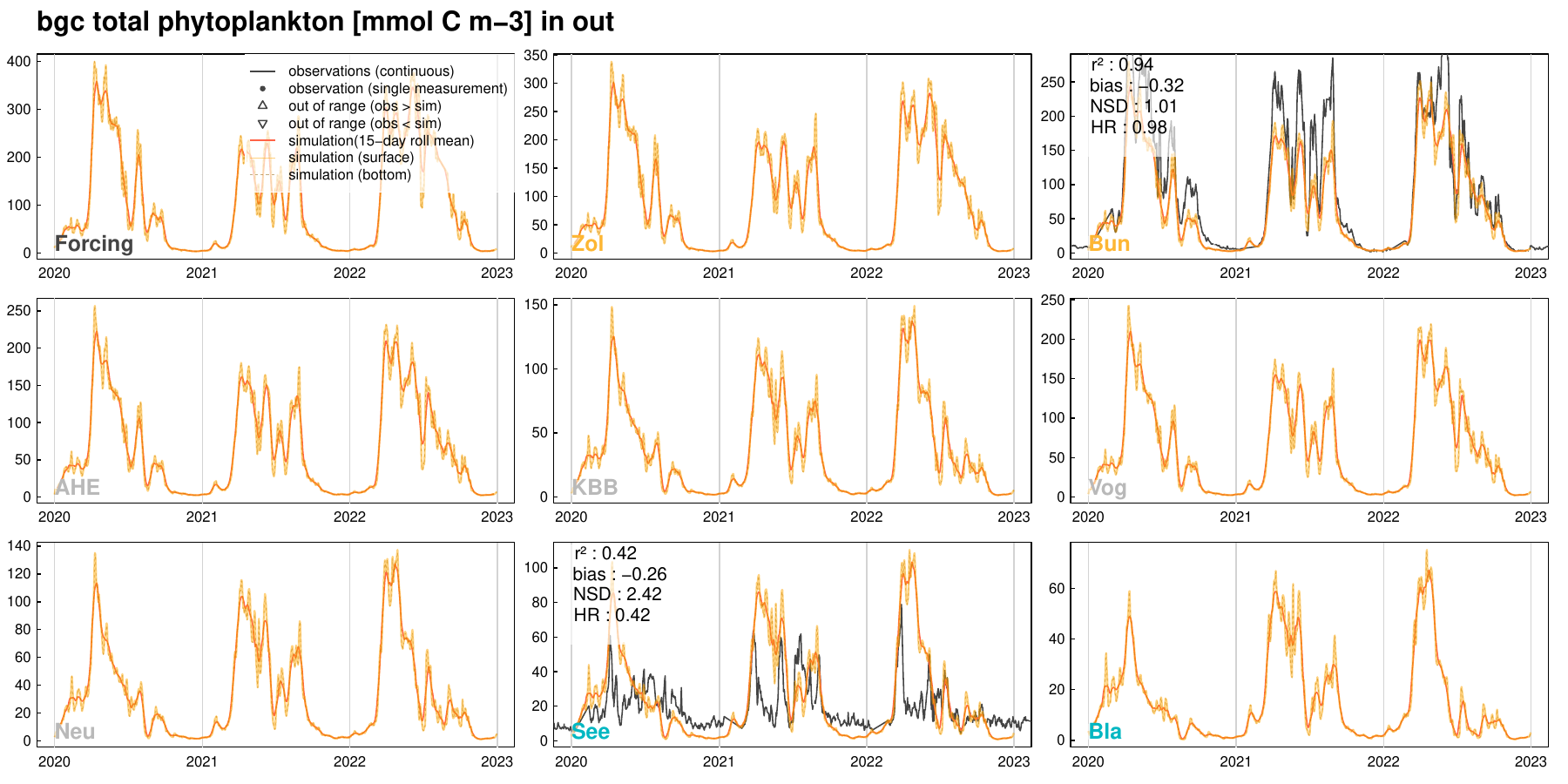} 
 \includegraphics[page=2,width=1.\textwidth]{sm-validation/out.bgc_ALG.validation} 
 \caption{\textbf{Total phytoplankton (total sum of both micro-algae classes) validation:} 
 Comparison of modelled (orange lines) and observed (black points and lines) values used as forcing in Geesthacht Wier and in each station (Zol = Zollenspieker, Bun = Bunthaus, AHE = Alte Harburger Elbbrücke, KBB = Köhlbrandbrücke, VNE = Vogelsander Norderelbe, Neu = Neumühlen, See = Seemannshöft, and Bla = Blankenese). 
 In each plot, we included the performance metrics for each station ($r^2$ = coefficient of determination, bias, NSD = normalized standard deviation, and HR = hit rate).
 The Taylor diagram (bottom left) and the aggregated observations vs. model comparisona and their performance metrics (bottom right) are also shown.}
 \label{fig:validation-alg}
\end{figure}

\begin{figure}[htb] %htb
 \centering
 \includegraphics[page=1,width=1.\textwidth]{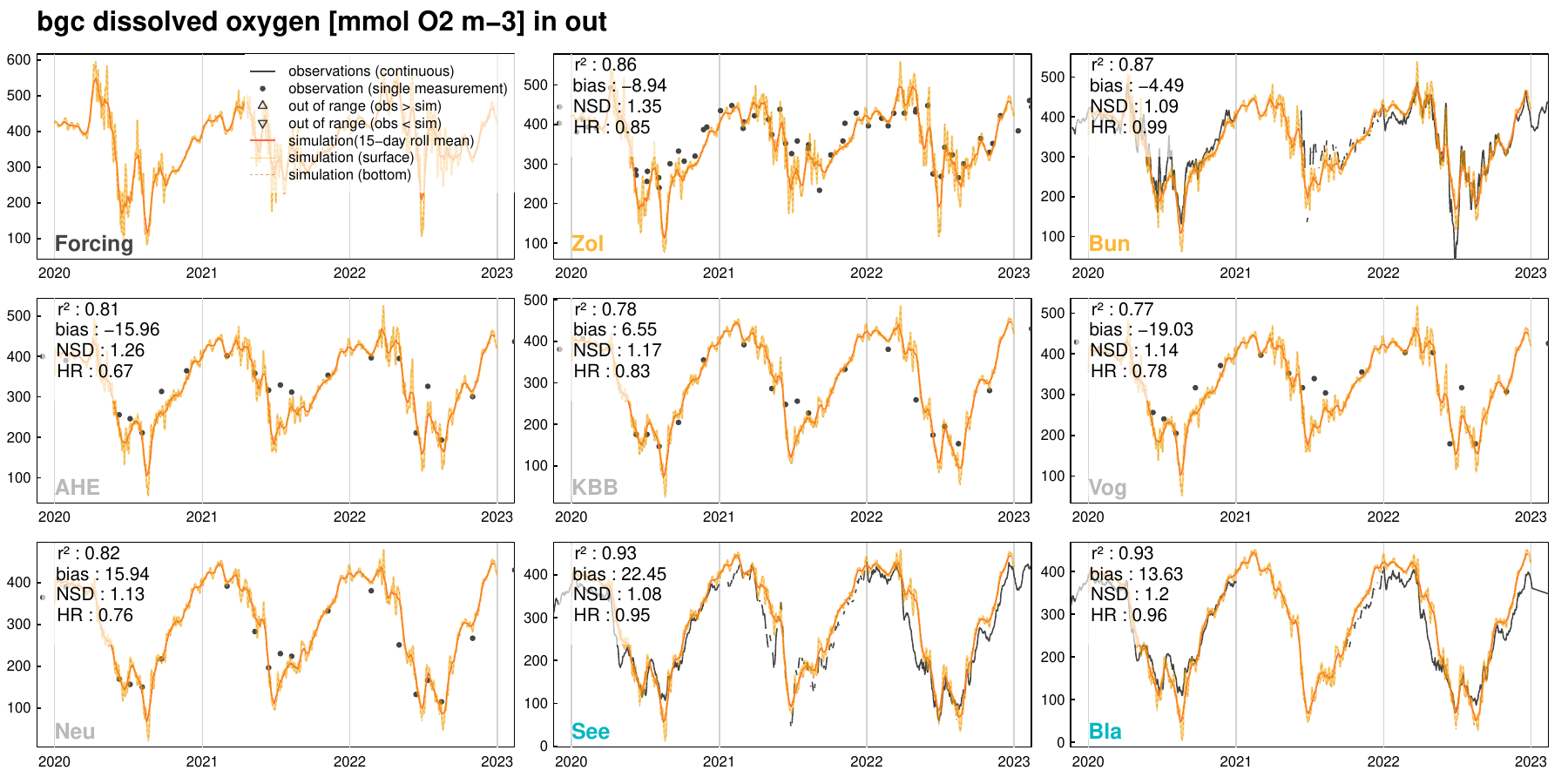} 
 \includegraphics[page=2,width=1.\textwidth]{sm-validation/out.bgc_DOxy.validation} 
 \caption{\textbf{Dissolved Oxygen validation:} 
 Comparison of modelled (orange lines) and observed (black points and lines) values used as forcing in Geesthacht Wier and in each station (Zol = Zollenspieker, Bun = Bunthaus, AHE = Alte Harburger Elbbrücke, KBB = Köhlbrandbrücke, VNE = Vogelsander Norderelbe, Neu = Neumühlen, See = Seemannshöft, and Bla = Blankenese). 
 In each plot, we included the performance metrics for each station ($r^2$ = coefficient of determination, bias, NSD = normalized standard deviation, and HR = hit rate).
 The Taylor diagram (bottom left) and the aggregated observations vs. model comparisona and their performance metrics (bottom right) are also shown.}
 \label{fig:validation-DO}
\end{figure}

\begin{figure}[htb] %htb
 \centering
 \includegraphics[page=1,width=1.\textwidth]{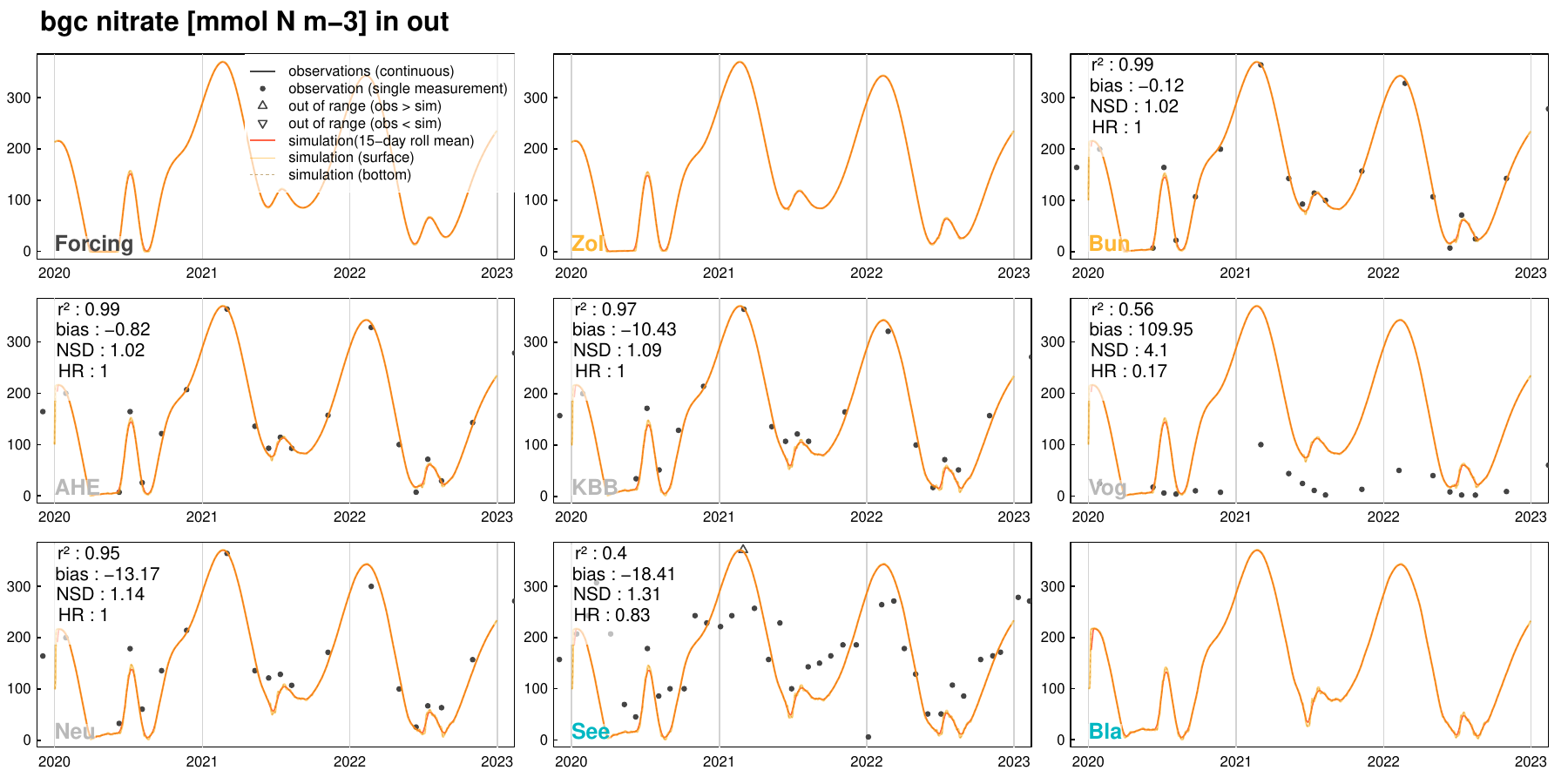} 
 \includegraphics[page=2,width=1.\textwidth]{sm-validation/out.bgc_NO3.validation} 
 \caption{\textbf{Nitrate validation:} 
 Comparison of modelled (orange lines) and observed (black points and lines) values used as forcing in Geesthacht Wier and in each station (Zol = Zollenspieker, Bun = Bunthaus, AHE = Alte Harburger Elbbrücke, KBB = Köhlbrandbrücke, VNE = Vogelsander Norderelbe, Neu = Neumühlen, See = Seemannshöft, and Bla = Blankenese). 
 In each plot, we included the performance metrics for each station ($r^2$ = coefficient of determination, bias, NSD = normalized standard deviation, and HR = hit rate).
 The Taylor diagram (bottom left) and the aggregated observations vs. model comparisona and their performance metrics (bottom right) are also shown.}
 \label{fig:validation-NO3}
\end{figure}

\begin{figure}[htb] %htb
 \centering
 \includegraphics[page=1,width=1.\textwidth]{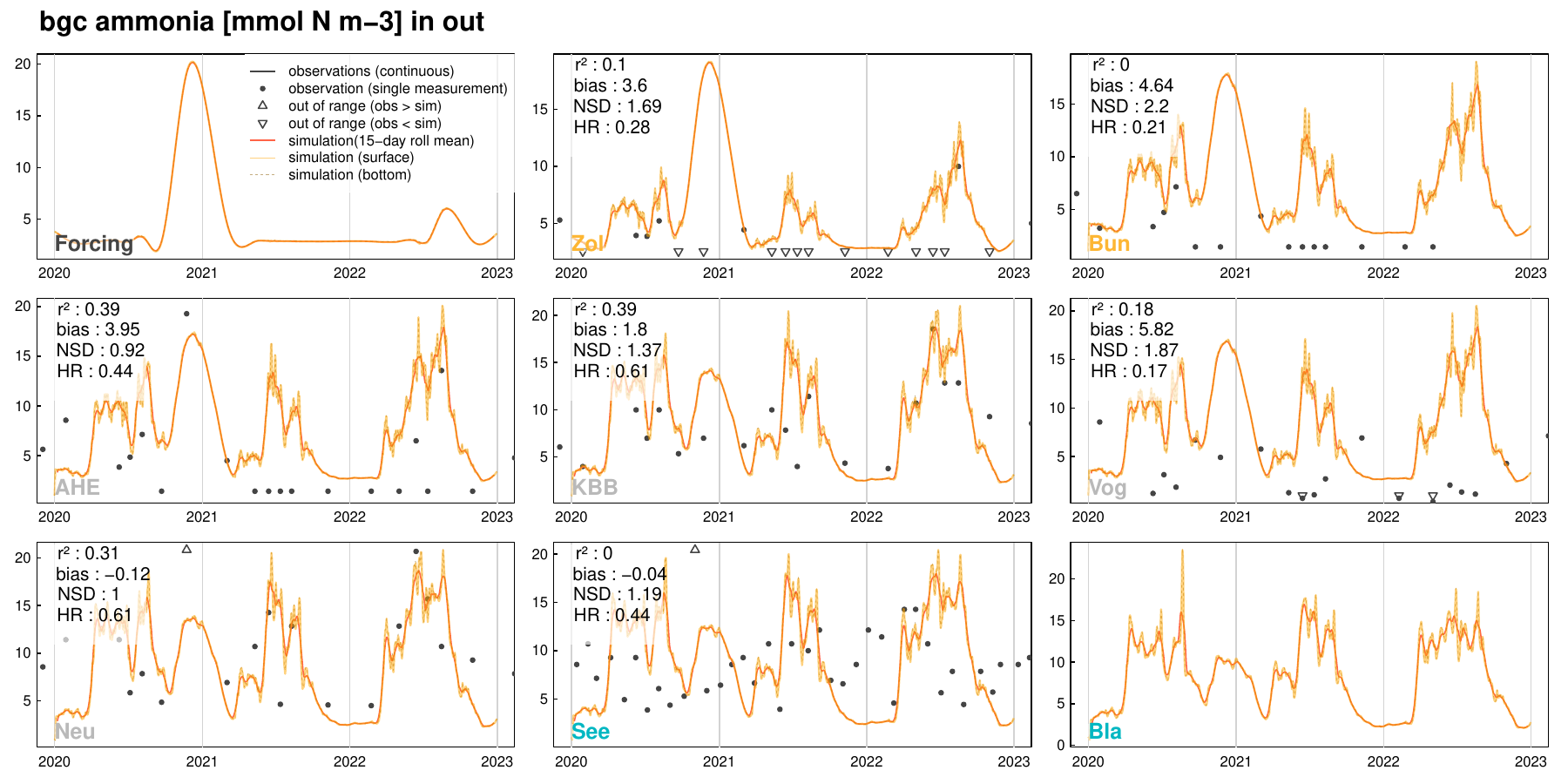} 
 \includegraphics[page=2,width=1.\textwidth]{sm-validation/out.bgc_NH4.validation} 
 \caption{\textbf{Ammonia validation:} 
 Comparison of modelled (orange lines) and observed (black points and lines) values used as forcing in Geesthacht Wier and in each station (Zol = Zollenspieker, Bun = Bunthaus, AHE = Alte Harburger Elbbrücke, KBB = Köhlbrandbrücke, VNE = Vogelsander Norderelbe, Neu = Neumühlen, See = Seemannshöft, and Bla = Blankenese). 
 In each plot, we included the performance metrics for each station ($r^2$ = coefficient of determination, bias, NSD = normalized standard deviation, and HR = hit rate).
 The Taylor diagram (bottom left) and the aggregated observations vs. model comparisona and their performance metrics (bottom right) are also shown.}
 \label{fig:validation-NH4}
\end{figure}

\begin{figure}[htb] %htb
 \centering
 \includegraphics[page=1,width=1.\textwidth]{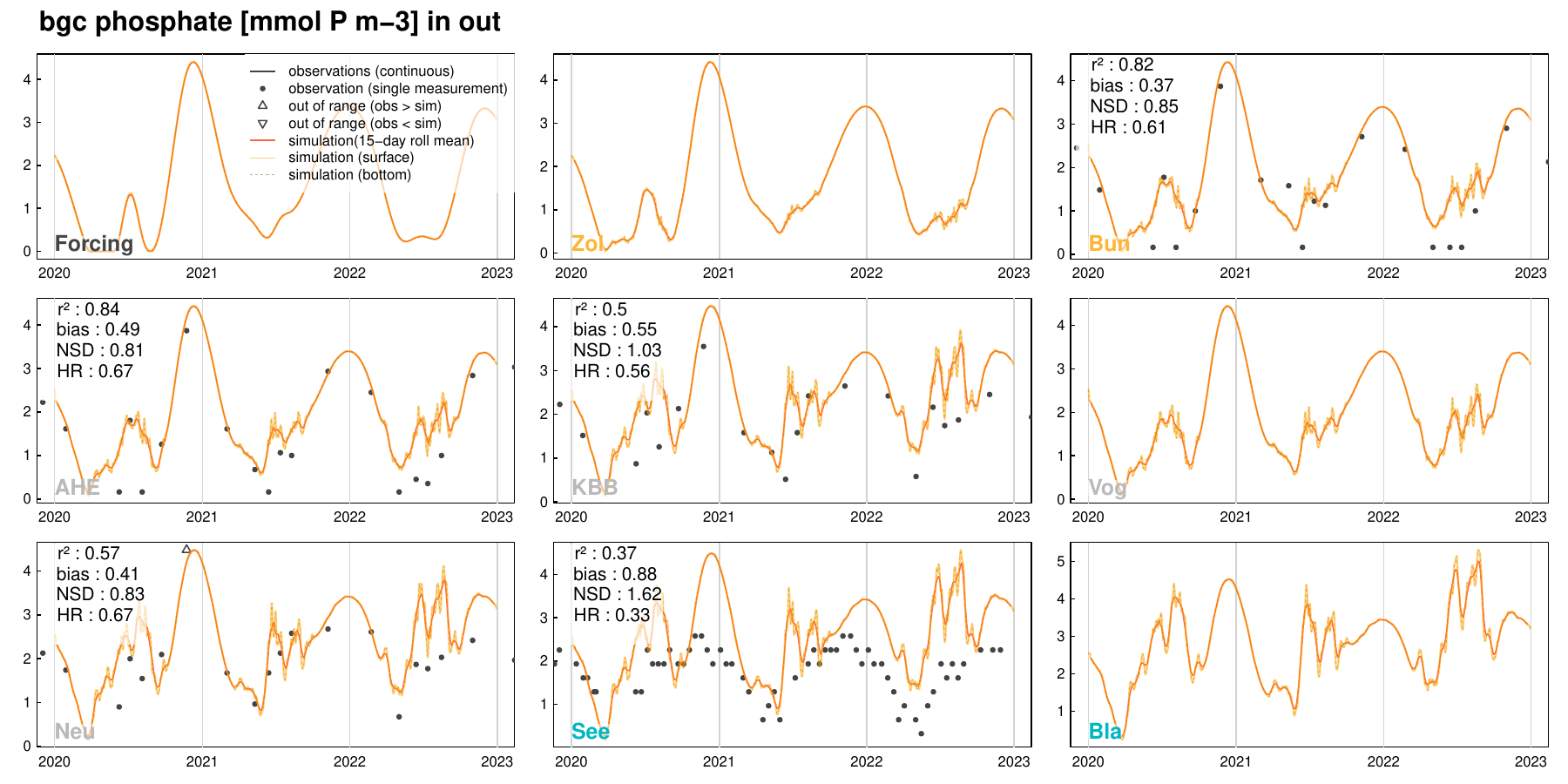} 
 \includegraphics[page=2,width=1.\textwidth]{sm-validation/out.bgc_PO4.validation} 
 \caption{\textbf{Ortho-phosphate validation:} 
 Comparison of modelled (orange lines) and observed (black points and lines) values used as forcing in Geesthacht Wier and in each station (Zol = Zollenspieker, Bun = Bunthaus, AHE = Alte Harburger Elbbrücke, KBB = Köhlbrandbrücke, VNE = Vogelsander Norderelbe, Neu = Neumühlen, See = Seemannshöft, and Bla = Blankenese). 
 In each plot, we included the performance metrics for each station ($r^2$ = coefficient of determination, bias, NSD = normalized standard deviation, and HR = hit rate).
 The Taylor diagram (bottom left) and the aggregated observations vs. model comparisona and their performance metrics (bottom right) are also shown.}
 \label{fig:validation-PO4}
\end{figure}

\begin{figure}[htb] %htb
 \centering
 \includegraphics[page=1,width=1.\textwidth]{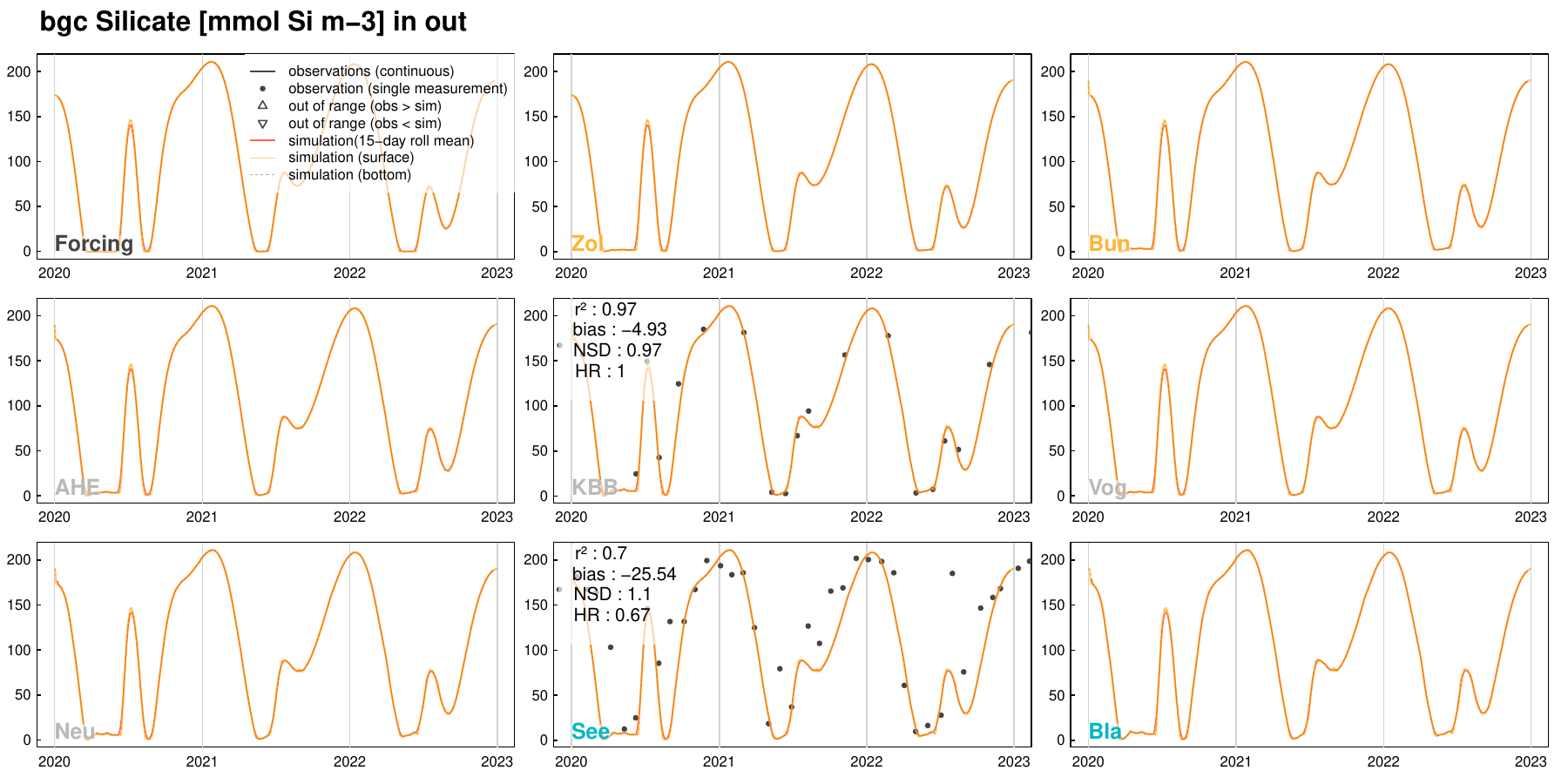} 
 \includegraphics[page=2,width=1.\textwidth]{sm-validation/out.bgc_Si.validation} 
 \caption{\textbf{Silica validation:} 
 Comparison of modelled (orange lines) and observed (black points and lines) values used as forcing in Geesthacht Wier and in each station (Zol = Zollenspieker, Bun = Bunthaus, AHE = Alte Harburger Elbbrücke, KBB = Köhlbrandbrücke, VNE = Vogelsander Norderelbe, Neu = Neumühlen, See = Seemannshöft, and Bla = Blankenese). 
 In each plot, we included the performance metrics for each station ($r^2$ = coefficient of determination, bias, NSD = normalized standard deviation, and HR = hit rate).
 The Taylor diagram (bottom left) and the aggregated observations vs. model comparisona and their performance metrics (bottom right) are also shown.}
 \label{fig:validation-Si}
\end{figure}

Our model application show high skill to reproduce observed values of 
temperature (Fig.~\ref{fig:validation-temp}),
total micro-algae concentration (Fig.~\ref{fig:validation-alg}),
dissolved oxygen (Fig.~\ref{fig:validation-DO}),
nitrate (Fig.~\ref{fig:validation-NO3}),
ammonia (Fig.~\ref{fig:validation-NH4}),
ortho-phosphate (Fig.~\ref{fig:validation-PO4}),
and silica (Fig.~\ref{fig:validation-Si}).

% Include here all the validation plots.

\end{document}